\documentclass[noshowpacs,onecolumn,showpacs,superscriptaddress,groupedaddress,amsmath,amssymb]{elsarticle}

\usepackage{amsmath}
\usepackage{amssymb}
\usepackage{amsthm}
\usepackage{xcolor}
\usepackage{amsbsy}
\usepackage{amstext}
\usepackage{graphicx}
\usepackage{color}
\usepackage{float}
\usepackage{siunitx}
\usepackage{wasysym}

\usepackage{mathtools}
\usepackage{subfigure}
\makeatletter

\usepackage{amsfonts}
\usepackage{dcolumn}
\usepackage{bbold,bm}

\usepackage{makecell}

\usepackage{hyperref}
\hypersetup{
    colorlinks=true, 
    linkcolor=blue, 
    urlcolor=red, 
    linktoc=all 
}

\usepackage[left=3cm,right=3cm]{geometry}

\def\bra#1{\mathinner{\langle{#1}|}}
\def\ket#1{\mathinner{|{#1}\rangle}}

\def\bbra#1{\mathinner{\langle\hspace{-0.75mm}\langle{#1}|}}
\def\kket#1{\mathinner{|{#1}\rangle\hspace{-0.75mm}\rangle}}

\def\dd{\mathrm{d}}
\def\ss{\mathbf{s}}

\def\dd{\boldsymbol{\delta}}

\newcommand{\vect}[1]{\boldsymbol{#1}}
\def\mnote#1{#1} 
\usepackage{accents}
\newlength{\hhatheight}
\newcommand{\hhat}[1]{%
    \settoheight{\hhatheight}{\ensuremath{\hat{#1}}}%
    \addtolength{\hhatheight}{-0.35ex}%
    \hat{\vphantom{\rule{1pt}{\hhatheight}}%
    \smash{\hat{#1}}}}

\makeatother

\begin{document}

\title{Low-frequency and Moir\'e Floquet engineering: a review}

\author[1,3]{Martin Rodriguez-Vega}
\ead{rodriguezvega@utexas.edu}
\author[2]{Michael Vogl}
\ead{ssss133@googlemail.com}
\author[3,4]{Gregory A. Fiete }

\address[1]{Department of Physics, The University of Texas at Austin, Austin, TX 78712, USA}
\address[2]{Department of Physics, King Fahd University of Petroleum and Minerals, 31261 Dhahran, Saudi Arabia}
\address[3]{Department of Physics, Northeastern University, Boston, MA 02115, USA}
\address[4]{Department of Physics, Massachusetts Institute of Technology, Cambridge, MA 02139, USA}

\begin{abstract}
We review recent work on low-frequency Floquet engineering and its application to quantum materials driven by light, emphasizing van der Waals systems hosting Moir\'e superlattices. These non-equilibrium systems combine the twist-angle sensitivity of the band structures with the flexibility of light drives. The frequency, amplitude, and polarization of light can be easily tuned in experimental setups, leading to platforms with on-demand properties. First, we review recent theoretical developments to derive effective Floquet Hamiltonians in different frequency regimes. We apply some of these theories to study twisted graphene and twisted transition metal dichalcogenide systems irradiated by light in free space and inside a waveguide. We study the changes induced in the quasienergies and steady-states, which can lead to topological transitions. Next, we consider van der Waals magnetic materials driven by low-frequency light pulses in resonance with the phonons. We discuss the phonon dynamics induced by the light and resulting magnetic transitions from a Floquet perspective. We finish by outlining new directions for Moir\'e-Floquet engineering in the low-frequency regime and their relevance for technological applications. 
\end{abstract}

\date{\today}
\maketitle
\tableofcontents

\section{Introduction}
\label{sec:introduction}

The term quantum material broadly refers to a condensed matter system where quantum mechanical effects manifest at the macroscopic level~\cite{Keimer2017, Tokura2017,Samarth2017}. These effects can originate from interactions or be rooted in the system's topology. For example, one of the oldest known magnetic materials, magnetite, is magnetic and conducts electricity at room temperature but becomes insulating at low temperatures due to complex charge ordering \cite{baldini2020,VERWEY,Senn2012}. On the opposite side of the spectrum, mercury is the oldest known superconductor \cite{onnes1911further} with vanishing resistivity at low-enough temperatures. In each case, the interactions of the internal degrees of freedom, such as electron spins, orbitals, lattice vibrations, etc., lead to ordered states. 

The quantum materials' properties can usually be tuned by changing external variables such as temperature, pressure, doping, chemical composition, and electromagnetic fields. A particularly striking example was discovered by the late Nobel laureate P. W. Anderson who showed that strong-enough random disorder could induce localized states~\cite{Anderson1958}. In later work, E. Abrahams, P. W. Anderson, et al.  introduced the concept of the Anderson insulator, a state which arises in non-interacting systems with weak random disorder. Anderson's ideas have led to generalizations including the effects on spin-orbit coupling~\cite{10.1143/PTP.63.707,PhysRevB.40.5325}, interactions~\cite{PhysRevB.26.7063,doi:10.1143/JPSJ.52.2870,PhysRevB.37.325,PhysRevB.58.R559}, and into the topological domain~\cite{Groth:prl09,LiJian2009,Hung:prb16,Chua:prb12,Prodan:prb11,Guo:prl10,Song:prb12,PhysRevB.93.125133}, including Floquet systems~\cite{PhysRevB.98.054203,PhysRevB.96.054207,titum2015,titum2016}. Anderson localization has been observed in several systems~\cite{Wiersma1997,Scheffold1999,Chabanov2000,Billy2008}.

In low-dimensional quantum materials, a rotation angle between layers provides a channel to control their states via the underlying moir\'e superlattice induced by the twist \cite{Bistritzer12233,dosSantos2012,morell2010}. For example, misaligned bilayer graphene on hexagonal boron nitride presents the fractal quantum Hall effect~\cite{Dean2013}. Twisted bilayer graphene (TBG) exhibits a series of superconducting, insulating~\cite{Cao2018,Cao2018sc,Codecidoeaaw9770,wong2019cascade,Lu2019efetov}, and ferromagnetic~\cite{Sharpe605, Seo_2019} states as a function of charge carrier (electron) concentration. Twisted double bilayer graphene (TDBG) displays spin-polarized and correlated phases~\cite{Cao2020,Shen2020,Liu2020,PhysRevB.99.235417,PhysRevB.99.235406,Lee2019,doi:10.1021/acs.nanolett.9b05117, C9NR10830K, PhysRevB.99.075127,kerelsky2019moireless,rubioverdu2020universal,halbertal2020moire}. There is evidence that twisted transition metal dichalcogenide (TTMD) heterostructures host moir\'e excitons~\cite{Tran2019,Seyler2019,Alexeev2019}. The manipulation of states of matter based on the twist angle is known as Moir\'e engineering, and provides yet another example where \textit{``More Is Different"}~\cite{Anderson393}. For recent reviews on Moir\'e heterostructures, see Refs.~\cite{kennes2020moire,andrei2020graphene}.
 
Since the ability to control the materials' properties can be exploited for their integration in new technologies, researchers have invested resources to elucidate new and efficient mechanisms. In the past decade, non-equilibrium approaches have emerged to control the plethora of correlated phases of matter on demand~\cite{Basov2017} and create novel topological phases~\cite{Oka_2019, rudner2020_review,Giovannini_2019,McIver2020,oka2009,lindner2011,rechtsman2013}. Next, we provide a brief overview of experimental results and theoretical proposals in periodically-driven condensed matter systems.

\begin{figure}[t]
	\begin{center}
		\includegraphics[width=8.0cm]{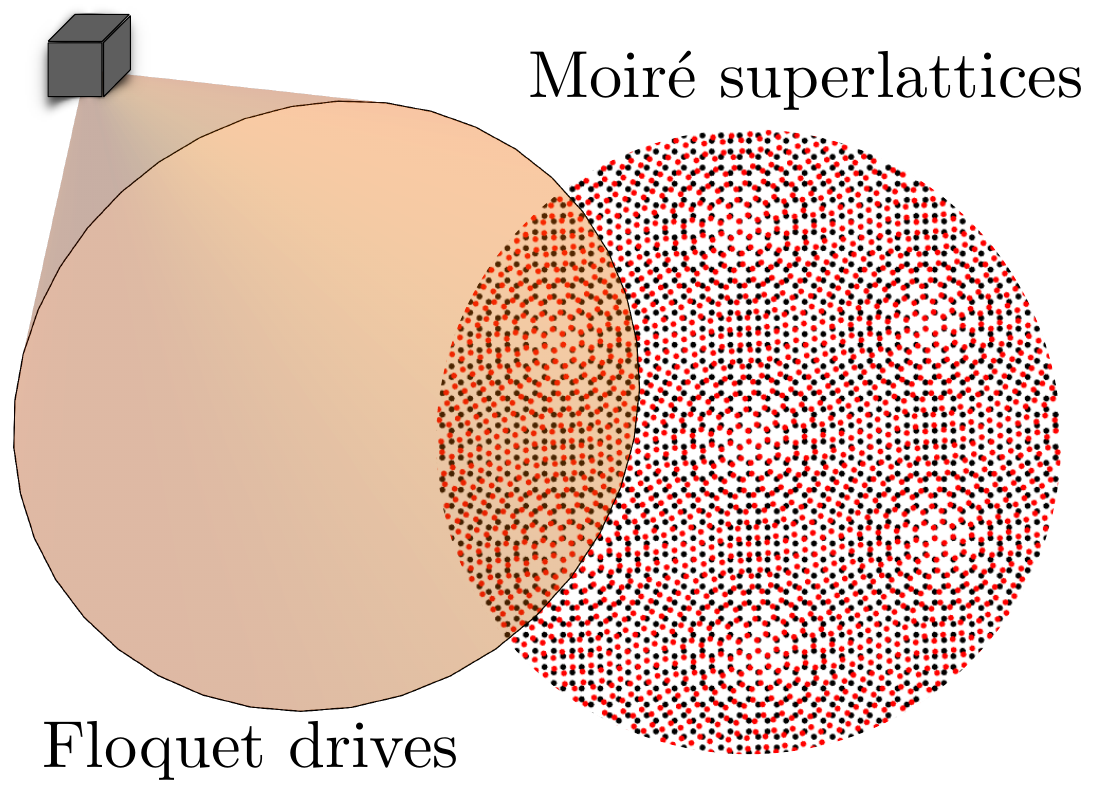}
		\caption{(Color online) Floquet-Moir\'e engineering describes the integrated use of light pulses at different frequencies to modify the properties of lattices with Moir\'e patterns. }
	\end{center}
	\label{fig:fpt_fig_main}
\end{figure}

Y. H. Wang \textit{et al. } reported the first observation of Floquet-Bloch States in Ref. \cite{Wang453}. They considered a strong topological insulator (Bi$_2$Se$_3$) and applied circularly polarized light with a frequency below the bulk band gap. Employing time- and angle-resolved photo-emission spectroscopy, they observed the quasienergy spectrum predicted by Floquet theory. Subsequent studies by F. Mahmood \textit{et al. } Ref. \cite{Mahmood2016} revealed scattering between Floquet–Bloch and Volkov states in Bi$_2$Se$_3$. In Ref.\cite{Sie2015}, E. J. Sie \textit{et al. } employed circularly polarized light to break time-reversal symmetry in transition-metal dichalcogenides to demonstrate valley-selective tunning of the exciton level in each valley. Several other phases have been experimentally realized. In particular, discrete-time crystals, which rely on many-body Anderson-localized states~\cite{Nandkishore2015,BASKO20061126}, have been predicted and realized\cite{Yao2017a,choi2017x,else2016,Khemani2016}. The light-induced anomalous Hall effect has been reported in graphene~\cite{McIver2020}, 
a photoinduced structural transition between semimetal phases have been observed in transition metal dichalcogenides via second-harmonic generation~\cite{Zhang2019-SymmetrySwitch}, and light-induced switching of stacking order in trilayer graphene~\cite{Zhang2020-trilayer}. For reviews on the theory and experimental aspects of photon-based spectroscopies, see Refs.~\cite{Wang2018-review} and ~\cite{Gedik2017}.

The theoretical proposals of Floquet phases are very extensive now. In particular, Anderson's localization theory has motivated topological generalizations in and out of equilibrium: in the topological Anderson insulator~\cite{LiJian2009}, a metal-to-insulator transition induced by disorder is accompanied by extended edge states. The disorder-induced Floquet topological insulators~\cite{titum2015} and the anomalous Floquet-Anderson insulator phase defined by non-adiabatic quantized charge pumping constitute extensions to non-equilibrium settings. Additional theoretical works include propositions for Kapitza pendulum-like many-body phases \cite{PhysRevB.100.104306}, anomalous Floquet states~\cite{Rudner2013} and their topological characterizations~\cite{PhysRevB.82.235114,PhysRevB.96.155118,PhysRevX.6.041001,PhysRevB.96.195303,Nathan_2015,zhang2020theory,doi:10.1146/annurev-conmatphys-031218-013721}, higher order Floquet topological phases~\cite{PhysRevLett.105.017401,PhysRevLett.123.016806,Huang_2020,PhysRevB.99.045441,PhysRevResearch.2.013124,PhysRevB.102.094305,chaudhary2019phononinduced,Haiping2020}, emergent Weyl semimetals and Fermi arcs~\cite{PhysRevB.96.041126,Zhu_2020,PhysRevLett.116.026805,Hubener2017,PhysRevLett.121.036401,PhysRevB.94.235137,PhysRevB.99.115136,PhysRevB.94.155206,PhysRevB.94.041409,PhysRevB.94.081103,PhysRevB.96.041205, PhysRevB.94.121106,PhysRevE.93.022209,PhysRevB.96.235424}, driven interacting systems~\cite{PhysRevLett.120.127601,PhysRevX.4.041048,PhysRevLett.118.115301,PhysRevB.98.045127,PhysRevB.93.245145}, driven superconductors~\cite{PhysRevB.98.214519,PhysRevB.101.054502,kundu2013,jiang2011,PhysRevB.100.035450,PhysRevB.98.235149,PhysRevB.96.155438,PhysRevLett.121.076802,PhysRevB.88.155133,benito2014} and light-induced d-wave superconductivity in cuprates~\cite{kennes2019}, metal-insulator transitions in semiconductors~\cite{Esineaay4922}, engineering of spin Hall insulators~\cite{PhysRevLett.108.056602}, topological frequency conversion\cite{MartinIvar2017}, Floquet transport and optical conductivity~\cite{PhysRevB.93.045121,PhysRevLett.115.106403,PhysRevB.97.245401,PhysRevB.96.165443,PhysRevLett.116.026801,PhysRevB.101.201401,PhysRevB.101.174314}, Floquet many-body localized phases~\cite{PhysRevLett.124.190601,PhysRevB.94.224202,Abanin_2016,PhysRevLett.114.140401,RevModPhys.91.021001}, and Floquet topological states in insulators~\cite{lindner2011,dallago2015,PhysRevB.91.241404,asboth2014}, graphene and hexagonal lattices~
\cite{calvo2011,zhenghao2011,PhysRevB.88.241112,perez2014,kundu2014,Sentef2015,Roman-Taboada2017,Dehghani2015,kitagawa2011,Usaj2014,PhysRevA.91.043625} and bilayer graphene~\cite{PhysRevB.86.125449,Qu_2017,dallago2017}. All these examples show how broad is the scope of Floquet systems.

The drive parameters such as the frequency, amplitude, phase or the shape of the drive provide additional knobs to modify the properties of the system. In particular, the frequency provides an energy scale relative to the quantum material. The non-equilibrium dynamics at large frequencies is well understood, within inverse-frequency expansions and rotating-wave approximation, as a renormalization of the equilibrium parameters of the system~\cite{blanes2009,rahav2003,rahav2003b,bukov2015,Eckardt_2015,Feldm1984,Magnus1954,PhysRevB.95.014112,PhysRevX.4.031027,PhysRevLett.115.075301,PhysRevB.93.144307,PhysRevB.94.235419,PhysRevLett.116.125301,Vogl_2019AnalogHJ,verdeny2013}. The mid- and low-frequency regimes, on the other hand, are relatively less explored~\cite{PhysRevLett.110.200403,Vogl2020_effham,vogl2019,Rodriguez_Vega_2018,PhysRevLett.121.036402,Martiskainen2015,rigolin2008,weinberg2015,PhysRevB.100.041103,PhysRevB.96.155438}. This regime is relevant for solid-state driven systems\cite{lindner2011} and provides a route to reduce heating effects when driving off electronic resonances~\cite{DAlessio2015,Weidinger2017,Str_ter_2016,PhysRevE.90.012110}. 

When frequencies in the terahertz range are employed to drive a system out of equilibrium, the phonons become relevant degrees of freedom. If strong electric fields are employed, the phonons are subjected to non-linear interactions, which can lead to an effective lattice control with implications for ordered states. Theoretically, the non-linear phononics mechanism has allowed the prediction of magnetic order tuning in RTiO$_3$ compounds ~\cite{kalsha2018,gu2018}, modulation of the structure of YBa$_2$Cu$_3$O, and related effects in the magnetic order~\cite{fechner2016}, light-enhanced superconductivity~\cite{Sentef2016}, and many other proposal~\cite{subedi2014, Mentink2015, gu2017, juraschek2017,Juraschek2017b,subedi2017,Babadi2017,liu2018,Juraschek2018,Juraschek2019,Juraschek2020,juraschek2019phonomagnetic,kalsha2018,gu2018,fechner2016,Sentef2016, Kennes2017, Sentef2017,Tancogne2018,chaudhary2019phononinduced,Chaudhary2019,vogl2020effective, vogl2020tuning, asmar2020floquet,chaudhary2020controlling, ke2020nonequilibrium, vogl2019,baldini2020,PhysRevLett.107.216601,PhysRevLett.116.016802,PhysRevMaterials.2.064401,PhysRevLett.125.137001}. Examples realized experimentally include the possibility to transiently enhance superconductivity~\cite{Fausti189,Mankowsky2014,mankowsky2015,mitrano2016}, light-induced metastable charge-density-wave states in 1T-TaS$_2$~\cite{Vaskivskyie1500168}, optical pulse-induced metastable metallic phases hidden in charge-ordered insulating phases ~\cite{Zhang2016, teitelbaum2019dynamics}, metastable ferroelectric phases in titanates~\cite{Nova1075}, photo-molecular superconductivity in charge-transfer salts\cite{PhysRevX.10.031028}, switching into meta-stable hidden phases~\cite{Stojchevska2014}, detection of displacive motion of Raman modes anharmonically coupled to driven infrared modes\cite{FORST201324}, and effective magnetic fields generated by multiple driven phonons~\cite{Nova2017}. For a review of experimental results on non-linear phononics, see Ref. \cite{Mankowsky_2016}. 

This review aims to provide an integrated picture of the recent developments of the Floquet theories valid the low-frequency regime, emphasizing their applications to Moir\'e superlattices.  The intersection of these subjects gives rise to Floquet - Moir\'e engineering of quantum materials. The rest of this review is organized as follows: In Sec. 2, we discuss several of the theories available to derive effective Floquet Hamiltonians valid in the high-, mid-, and low-frequency regimes. We provide an overview of the advantages and drawbacks of each of the methods discussed. We use this section to introduce the relevant definitions and highlight the technical challenges in the low-frequency regime, the main focus of this work. In Sec. 3, we apply the methods previously discussed to analyze quantum materials driven by light in the mid-, and low-frequency regimes. Finally, in Sec. 4 we conclude with a summary and our perspectives and outlook for low-frequency Floquet physics.

\section{Effective theories}

This section will discuss various methods to obtain effective or approximate Floquet Hamiltonians to describe periodically-driven systems' stroboscopic dynamics. Our focus will be on the theories that are, in a sense, averaged over a single period, and we will therefore ignore the issue of micromotion operators. First, we introduce two different pictures used to describe Floquet systems: real-time and frequency domains. Then, we review some of the theories available in the high-, mid- and low-frequency regimes.

\subsection{Two paths to Floquet theory}

There are two common starting points for describing Floquet systems, both with their advantages and disadvantages. In the first part, we will offer short derivations of both and explain their use and their insights. This will also serve us well in later sections when we explain techniques based on both approaches.

\subsubsection{From the perspective of a time evolution operator}

In the first common approach \cite{PMID:21806092,https://doi.org/10.1002/cmr.a.21414,blanes2009,bukov2015,KUWAHARA201696,Feldm1984,PhysRevB.25.6622,PhysRevLett.116.120401,vogl2019,Vogl_2019AnalogHJ} 
one may take the Schr\"odinger equation for the time evolution operator $U(t)$
\begin{equation}
	i\partial_t U(t)=H(t)U(t)
\end{equation}
as a starting point.
A formal solution to this equation is given by
\begin{equation}
	U(t)=\mathcal{T}\exp(-i\int dt H(t)).
\end{equation}
This object is called a time ordered exponential and can be defined as a product of operator exponentials as
\begin{equation}
	U(t)=\lim_{\delta t\to 0} \prod_{n=1}^{t/\delta t} e^{-i dt H(n \delta t)}.
	\label{produc_time_ord_exp}
\end{equation}
If the Hamiltonian is periodic in time $H(t)=H(t+T)$ then the problem has an important simplification. Namely Eq. \eqref{produc_time_ord_exp} then allows us to read off that
\begin{equation}
	U(nT)=U(T)^n.
\end{equation}
This allows us to interpret the equation as $U(nT)=\exp(-iH_F nT)$ and to define the so-called Floquet Hamiltonian
\begin{equation}
	H_F:=i\log(U(T))/T
\end{equation}
that determines the time evolution at stroboscopic times $t=nT$ and behaves similar to the Hamiltonian of a time-independent system in this respect.

Let us first recognize that this definition includes a matrix logarithm with many branches and, therefore, multiple possible solutions that are equivalent in their time evolution properties. Next, we interpret this definition physically. If one is only interested in what happens at so-called stroboscopic times $nT$ then one can straightforwardly see that $H_F$ acts in the same way that a time-independent Hamiltonian would. In this sense, one may interpret the quantity of $H_F$ as an effective time-independent Hamiltonian. If $T$ is sufficiently small, one may replace $nT\to t$ and $H_F$ will describe dynamics that are smoothed out in time and represented by $U(t)\approx e^{-iH_Ft}$ just like a time-independent system. One should note that such a description in some cases may even be more advantageous in describing experiments than the fully detailed description with better time resolution. Afterall detectors often implicitly average over short time frames, and therefore a coarse-grained description such as by $H_F$ may be more easily related to some experiments. Regardless one may not just interpret $H_F$ as a generator of time evolutions but also use it to predict phases of matter without taking recourse to the complicated methods of time-dependent quantum mechanics. This is because it describes states in an exponentially long time regime called the pre-thermal regime \cite{PhysRevB.95.014112}.  One should notice that $H_F$ is implicitly dependent on the initial conditions that are chosen since those enter a time-ordered exponential. However, often this is of no vital consequence.

The main technical challenges for this approach are: (i) computing the time-ordered exponential $U(T)$ and (ii) taking the matrix logarithm to find $H_F=i \log(U(T))/T$. In later sections, we will discuss some approaches that make it possible to simplify these problems.

\subsubsection{From the perspective of a quasi-energy operator}
\label{sec:From the perspective of a quasi-energy operator}

In a second common approach \cite{Eckardt_2015,PhysRevB.93.144307,PhysRevA.7.2203,PhysRevLett.111.175301,Rodriguez_Vega_2018,Vogl2020_effham} one may consider the time dependent Schr\"odinger equation for the wavefunction
\begin{equation}
	i\partial_t\psi=H(t)\psi,
\end{equation}
as a starting point for Floquet theory.  This equation (analogous to the discussion in the previous section), can be simplified if the Hamiltonian is periodic in time, $H(t+T)=H(t)$, because the Hamiltonian then commutes with the stroboscopic shift operator $\hat T H(t)=H(t+T)$, that is $[H(t),\hat T]=0$. This means that one can find simultaneous eigenvalues of $\hat T$ and $H(t)$, which will allow a block-diagonalization if we can diagonalize $\hat T$.

Our first step therefore is to construct $\hat T$. To do so we study the wavefunction at times $t+T$. Via a Taylor expansion we find
\begin{equation}
	\psi(t+T)=e^{i(-iT\partial_t)}\psi(t)=\hat T\psi(t).
	\label{psitpT}
\end{equation}
We immediately find the stroboscopic shift operator as $\hat T=e^{i(-iT\partial_t)}$. Next we find eigenvalues of $\hat T$ as
\begin{equation}
	\hat T\phi=e^{i\epsilon T} \phi.
\end{equation}

This allows us to separate the wavefunction according to $\psi(t) =e^{i\epsilon t} u(t)$ and we find that
\begin{equation}
	u(t+T)=e^{-i\epsilon(t+T)}\psi(t+T)=e^{-i\epsilon t}\psi(t)=u(t)
\end{equation}
if we use the eigenvalue in Eq.\eqref{psitpT}, rearrange and multiply by $e^{-i\epsilon t}$. Therefore, we are able to separate the wavefunction into a periodic part and a phase. Finally, it is clear that with the ansatz $\psi(t) =e^{i\epsilon t} u(t)$ we are able to block-diagonalize as
\begin{equation}
	\epsilon u(t)=(H(t)-i\partial_t)u(t)=\hat Qu(t),
	\label{quasieneq}
\end{equation}
where the equation separates into different blocks labeled by the quantity $\epsilon$, and $t$ takes values in a Floquet zone, $[t_0,t_0+T]$ with arbitrary $t_0$, instead of the full real axis, similar Bloch's theorem.

Thus, Eq.\eqref{quasieneq} is now an eigenvalue equation for a time-periodic wavefunction $u(t)$ that has a similar structure to the time-independent Schr\"odinger equation. However, the quantity $\epsilon$ is not the energy but quasi energy, and $\hat Q=H(t)-i\partial_t$ is not the Hamiltonian but a quasi energy operator that acts in a larger Hilbert space. More precisely, we notice that time enters in a new way into the equation--it takes a footing similar to positions $x$ in the ordinary Schr\"odinger case. We stress that while the structure is the same as equilibrium quantum mechanics, the operator $\hat Q$ lives in an extended Hilbert space $F=H\otimes I$, where $H$ denotes the original Hilbert space and $I$ the space of functions on the time interval of one period $T$. Therefore, the scalar product from the original Hilbert space is not sufficient since it does not include contributions from the time variable and one needs to define a new scalar product. Generically, a physically meaningful scalar product for vectors $u$ and $v$ is usually defined by a sum over all indices that can label states like $\langle u|v\rangle= \sum_i u_i^* v_i$ or in the case of continuous labels by an integral. In our case, the new variable time takes continuous values. One may now build additional structure onto the scalar product for the original Hilbert space $\langle .,.\rangle$ and define a new scalar product as
\begin{equation}
	\langle\hspace{-0.75mm}\langle u, v\rangle\hspace{-0.75mm}\rangle=\int_0^T dt  \langle u(t),v(t)\rangle,
\end{equation}
which is just a sum over all labels in the extended Hilbert space.  Since the systems of interest have period $T$ it would be redundant to integrate over longer times.

It is also useful to define a partial scalar product that acts on time labels 
\begin{equation}
    ( u, v)=\int_0^T dt u(t)v(t)
\end{equation}
as we will see shortly.  This way of framing the problem in an extended space is not immediately useful but similar to the Bloch problem it can be cast into a form that will be useful for numerical computations.

One may use the partial scalar product $( ., .)$ to express $\hat Q$ in a Fourier basis $|n ) =e^{in\omega t}/\sqrt{T}$. After a short computation one may therefore recast the quasi-energy Eq.\eqref{quasieneq} as
\begin{align}
&\epsilon \begin{pmatrix}
\vdots\\
u_{-1}\\
u_0\\
u_1\\
\vdots
\end{pmatrix}=\begin{pmatrix}
\ddots&\vdots&\vdots&\vdots&\vdots&\vdots&\\
\cdots&H_1& H_0-\omega&H_{-1}&H_{-2}&H_{-3}&\cdots\\
\cdots&H_2&H_1&  H_0&H_{-1}&H_{-2}&\cdots\\
\cdots&H_3&H_2&H_1& H_0+\omega&H_{-1}&\cdots\\
&\vdots&\vdots&\vdots&\vdots&\vdots& \ddots
\end{pmatrix}\begin{pmatrix}
\vdots\\
u_{-1}\\
u_0\\
u_1\\
\vdots
\end{pmatrix},
\label{quasi-en-eq}
\end{align}
where $u(t)=\sum_n e^{in\omega t} u_n$.  The different entries in this matrix are defined as $H_n=( m+n| H(t)|m)=\frac{1}{T}\int_0^T dt e^{-i n\omega t} H(t)$, which are Fourier components of the time-dependent Hamiltonian. 

The advantage of this approach over that from the previous section is that it is independent of the initial conditions. Of course, one has to note that the expression in Eq.\eqref{quasi-en-eq} is only useful in the case that it is truncated, which often is an excellent approximation and is extensively used in numerical studies. The disadvantage of this approach compared to the one from the previous section is that it is less simple to access the time evolution operator. However, one is still able to construct full time dependent wavefunctions by recourse to $u(t)=\sum_n e^{in\omega t} u_n$ and Eq.\eqref{quasi-en-eq}. 

It is also important to note that the quasi energy $\epsilon$ is an experimentally observable quantity. In the limit $\omega\to \infty$, Eq.\eqref{quasi-en-eq} predicts copies of the spectrum of $H_0$ that are shifted by $\omega$ if $H_n$ is negligible for $|n|>0$. These Floquet copies have been observed in experiments\cite{wang2013,Mahmood2016}.

\subsection{High-frequency approximations}

After surveying the two common approaches to describe the Floquet theory, we will now see how we can obtain useful approximations. For this, we will start with the high frequency regime, characterized by driving frequency $\Omega\gg h$ where $h$ is the magnitude of local terms in a Hamiltonian.

\subsubsection{Floquet-Magnus expansion}

We first review the Floquet-Magnus expansion, based on the time evolution operator. One of the difficulties of computing the Floquet operator $H_F$ is that we have to compute a matrix logarithm. Magnus \cite{original_Magnus_paper} side-stepped this issue by seeking a solution for $U(t)$ of the form
\begin{equation}
    U(t)=e^{\Omega(t)},
    \label{Magnus_ansatz}
\end{equation}
which makes it possible to read off $H_F=i\Omega (T)/T$~\cite{https://doi.org/10.1002/cmr.a.21414,bukov2015,KUWAHARA201696,Feldm1984,PhysRevB.25.6622,PhysRevLett.116.120401,Vogl_2019AnalogHJ}. The main problem is now finding an expression for $\Omega(t)$. In principle this can be done as follows. If we insert Eq.\eqref{Magnus_ansatz} into the Schr\"odinger equation for the time evolution operator we find that $\Omega(t)$ has to fulfill
\begin{equation}
    i\partial_t \Omega(t)=\frac{ad_{\Omega(t)}}{\exp(ad_{\Omega(t)})-1}H(t),
    \label{eq_Magnus_generator}
\end{equation}
where we have used the derivative of the exponential map \begin{equation}
    \frac{d}{dt}e^{X(t)}=e^{X(t)}\frac{1-\exp(-ad_{X(t)})}{ad_{X(t)}}\frac{d X(t)}{dt}
\end{equation}
and the short-hand notation for the adjoint map $ad_A=[A,.]$ and therefore $ad_{A}^2=[A,[A,.]]$ etc.

At first glance, we see that Eq.\eqref{eq_Magnus_generator} is more complicated than the Schr\"odinger equation. However, this reformulation is well-suited to construct $\Omega(t)$ perturbatively in powers of $H(t)$ by starting with the lowest order approximation $\Omega_0(t)=0$ and finding higher corrections in a series $\Omega(t)=\sum_n \Omega_n(t)$. The first two terms are
\begin{equation}
    \Omega_1(t)=-i\int_0^t dt_1 H(t_1);\quad \Omega_2(t)=-\int_0^t dt_1\int_0^{t_1} dt_2[H(t_1),H(t_2)].
    \label{magnus_ord_2}
\end{equation}
Therefore, one may directly find an expression for the Floquet Hamiltonian as $H_F=\sum_n i\Omega_n(T)/T=\sum_n H_{FM}^{(n)}$ and $H_{FM}^{(n)}= i\Omega_n(T)/T$. One should note that each term $H_{FM}^{(n)}$ is of order $\omega^{-n}$. This can be seen by recognizing that each term will have an additional integral. More precisely, for a periodic Hamiltonian $H(t)=H(t+T)$ and $T=2\pi/\omega$. One can introduce a new integration variable $\tau=\omega t$ and then finds that each term $dt=(1/\omega)d\tau$. It is now clear why this is considered a high frequency expansion.

To ensure that the results can be compared with subsequent sections we note that that the expansion can also be written in terms of Fourier components of the Hamiltonian $H_n=\frac{1}{T}\int_0^T dt e^{-i n\omega t} H(t)$ as \cite{PhysRevB.93.144307}
\begin{equation}
H_{FM}^{(0)}=H_0;\quad H_{FM}^{(1)}=\sum_{n\neq 0}\frac{[H_{-n}-2H_0,H_n]}{2\omega n}.
\end{equation}
In this form it is not obvious that there is a dependence on initial conditions. This can be made explicit if instead of $t=0$  as starting point for integrations in \eqref{magnus_ord_2} we choose $t=t_0$ to find
\begin{equation}
H_{FM}^{(1)}=\sum_{n\neq 0}\frac{[H_{-n}-2H_0e^{-in\omega t_0},H_n]}{2\omega n}.
\end{equation}

Now the effective Hamiltonian $H_{FM}^{(1)}$ has an explicit dependence on the initial time. The choice of $t_0$ is defined as the Floquet gauge~\cite{bukov2015}.

\subsubsection{Van Vleck expansion}

We next want to review one of the most important methods to determine a perturbative expansion for an effective Hamiltonian  independent of an arbitrary initial time $t_0$. For this, the extended-space picture, Eq.\eqref{quasi-en-eq}, is a good starting point since it does not explicitly depend on time. Following the discussion in Ref.\cite{Eckardt_2015}, we can find an effective Hamiltonian if we block-diagonalize the quasi-energy operator Eq.\eqref{quasi-en-eq} in photon-number space (the indices in the equation). This can be achieved by a yet undetermined unitary transformation $U=e^{G}$, where $G^\dag=-G$ is anti-hermitian. 

Let us first we rewrite the quasienergy operator using a dummy parameter $\lambda=1$ as
\begin{equation}
    \hat Q=\lambda (\mathcal{H}_D+\mathcal{H}_X)+\omega\mathcal{M},
\end{equation}
where $[\mathcal{M}]_{mn}=m\delta_{mn}$ describes the unperturbed problem, $[\mathcal{H}_D]_{mn}=\delta_{mn}H_{m-n}$ the diagonal (in photon space) part of the perturbation, $[\mathcal{H}_X]_{mn}=(1-\delta_{mn})H_{m-n}$ the offdiagonal part and
$H_{n}=\frac{1}{T}\int_0^T dt e^{-in\omega t}H(t)$.

If we want to block-diagonalize the quasienergy operator via the unitary transformation $e^{G}$ then the diagonal (signified by a subscript D) and off-diagonal parts (signified by a subscript X) have to fulfill
\begin{equation}
[e^{-G} (\lambda (\mathcal{H}_D+\mathcal{H}_X)+\mathcal{M})e^{G}]_D=\omega \mathcal{M}+W
\end{equation}
 and 
 \begin{equation}
 [e^{-G} (\lambda (\mathcal{H}_D+\mathcal{H}_X)+\mathcal{M})e^{G}]_X=0,
 \end{equation}
where $W$ is a yet undetermined matrix that is related to the effective Hamiltonian via $W_{nn}=H_{\mathrm{eff}}$.

Since we have two equations we are now able to determine both $W$ and $G$. We will do so by means of a power series ansatz $G=\sum_{n=1}^\infty\lambda G^{(n)}$ and $W=\sum_{n=1}^\infty \lambda^n W^{(n)}$. We will also assume that $G$ is off-diagonal, that is $G=G_X$.  Once we collect different orders of $\lambda$, we find for the lowest few orders that
\begin{equation}
     \begin{aligned}
    &[G^{(1)},\omega \mathcal{M}]=\mathcal{H}_X;\quad [G^{(2)},\omega \mathcal{M}]=[\mathcal{H}_D,G^{(1)}]+\frac{1}{2}[\mathcal{H}_X,G^{(1)}]_X\\
    &W^{(1)}=\mathcal H_D;\quad W^{(2)}=\frac{1}{2}[\mathcal{H}_X,G^{(1)}]_D\\
    &W^{(3)}=\frac{1}{2}[\mathcal H_X,G^{(2)}]_D+\frac{1}{12}[[\mathcal H_X,G^{(1)}],G^{(1)}]_D.
    \end{aligned}
\end{equation}
These equations can be solved for the different blocks of $G^{(n)}$ (in photon-number space)  and subsequently $W^{(n)}$ if one recognizes that
\begin{equation}
   [G,\mathcal{M}]_{ij}=(j-i)G_{ij}.
\end{equation}
 Making use of this identity and recalling $W_{0,0}=H_{\mathrm{eff}}$ one can find an effective Hamiltonian $H_{\mathrm{eff}}=\sum_{n=1}^\infty H^{(n)}_{vV}$ with the first two orders given as 
 \begin{equation}
 \begin{aligned}
  &H^{(1)}_{vV}=H_0,\\
  &H^{(2)}_{vV}=\sum_{m\neq 0}\frac{H_mH_{-m}}{m\omega},\\
  &H^{(3)}_{vV}=\sum_{m\neq0}\left(\frac{[H_{-m},[H_0,H_m]]}{2m^2\omega^2}+\sum_{m\neq0,m}\frac{[H_{-m},[H_{m^\prime-m},H_m]]}{3mm^\prime \omega^2}\right).
 \end{aligned}
 \end{equation}
A detailed discussion is included in Refs. \cite{Eckardt_2015,PhysRevB.93.144307}. The advantage of this method lies in that the effective Hamiltonian does not depend on initial conditions $t_0$ and it keeps symmetries explicitly manifest that may be broken in the case of the Magnus expansion \cite{Eckardt_2015}.

\subsubsection{Brillouin Wigner expansion}

We next review another alternative high frequency expansion that also does not depend on an initial time $t_0$ but will have more compact expression for higher order effective Hamiltonian than the van Vleck expansion. For this we follow Ref.\cite{PhysRevB.93.144307} and again choose Eq. \eqref{quasi-en-eq} as our starting point, which in this section we write as
\begin{equation}
   \hat Q\psi= (\mathcal{H}-\mathcal{M}\omega)\psi=\epsilon\psi,
   \label{quasi_en_eq_BW}
\end{equation}
where $[\mathcal{M}]_{mn}=m\delta_{mn}$ and $[\mathcal{H}]_{mn}=H_{mn}=\frac{1}{T}\int_0^T dt e^{i(m-n)\omega t}H(t)$.

Since we will only be interested in one block for the effective Hamiltonian we want to write this equation projected on a subspace defined by a projection operator $\mathcal{P}$. To achieve this, one can just project the full operator equation as
\begin{equation}
\mathcal{P}\hat Q\psi=\epsilon \mathcal{P}\psi,
\label{projected}
\end{equation}
where $\psi_P=\mathcal{P}\psi$ is the projected eigenvector. Equation \eqref{projected} is not an eigenvalue equation for $\psi_P$ yet. However, we can achieve this if we introduce the wave-operator $\hat \Omega$ that allows us to disassemble the identity as $\mathbb{1}=\hat \Omega \mathcal{P}$. To gain intuition about the wave-operator $\hat \Omega$ it is important to note that $\psi=\hat \Omega \psi_P$, which tells us that it includes the information to reconstruct full-space eigenvectors from the projected eigenvectors. From here we can directly see that Eq. \eqref{projected} can be rewritten as an eigenvalue equation for the projected eigenvector $\psi_P$ as
\begin{equation}
H_{\mathrm{eff}}\psi_P=\epsilon \psi_P;\quad H_{\mathrm{eff}}=\mathcal{P}\hat Q\hat\Omega \mathcal{P}.
\end{equation}

Now that we have a formal definition of the effective quasienergy operator in an arbitrary projected space, we can choose what space we are interested in. We do so by choosing an appropriate projection operator $\mathcal P$ and have to find a corresponding wave operator $\hat \Omega$. For the high frequency approximation we choose the sub-space with zero photons that is described by the projection operator $[\mathcal{P}]_{mn}=\delta_{mn}\delta_{m0}$. 

The wave-operator $\hat \Omega$ can now also be constructed in the following fashion. We first project Eq. \eqref{quasi_en_eq_BW} with the complementary projection operator $\bar{\mathcal P}=\mathbb{1}-\mathcal P$ to find the auxiliary identity
\begin{equation}
    \bar{\mathcal P}\psi=\frac{\bar{\mathcal P}}{\epsilon+m\mathcal{M}}\mathcal{H}\psi.
    \label{BW_Auxiliary}
\end{equation}
Next we decompose $\psi=\psi_P+\bar{\mathcal P}\psi$, which by using Eq.\eqref{BW_Auxiliary} can be rearranged into
\begin{equation}
\psi=\hat \Omega \psi_P;\quad \hat \Omega=\left(\mathbb{1}-\frac{\bar{\mathcal P}}{\epsilon+\mathcal{M}\omega}\mathcal{H}\right)^{-1},
\end{equation}
and we have therefore found an expression for the wave-operator. Finally we are now able to compute the exact effective Hamiltonian explicitly as
\begin{equation}
H_{\mathrm{eff}}=\mathcal{P}\mathcal{H}\left(\mathbb{1}-\frac{\bar{\mathcal P}}{\epsilon+\mathcal{M}\omega}\mathcal{H}\right)^{-1} \mathcal{P},
\end{equation}
where we have used $\mathcal P\mathcal M=0$. Explicitly, this can also be written as (for instance, if we use Eq. \eqref{BW_Auxiliary} in $\psi=\psi_P+\bar{\mathcal P}\psi$ and iterate) 
\begin{equation}
    H_{\mathrm{eff}}=H_{00}+\sum_{N=1}^\infty\sum_{n_1,...,n_N\neq 0}\frac{H_{0,n_1}(\prod_{i=1}^{N-1} H_{n_i,n_i+1})H_{n_N,0}}{\prod_{i=1}^N(\epsilon+n_i\omega)}.
    \label{eq:BW_gen}
\end{equation}

The Eq.\eqref{eq:BW_gen} appears to have the disadvantage that it explicitly depends on quasi-energy $\epsilon$ and therefore would lead to a Schr\"odinger equation that would have to be solved self-consistently with respect to $\epsilon$. As we will discuss in Sec. \ref{sec:selfconsistentlow}, this self-consistent condition is inherent of the low-frequency regime. It is needed to capture non-analytic behaviour which induces bandgap openings in the quasienergies in this regime.

The self-consistent condition is lifted if we expand for large $\omega$. We find $H_{\mathrm{eff}}=\sum_{n=0}^\infty H_{BW}^{(n)}$ with the first few terms given as
\begin{equation}
    H_{BW}^{(0)}=H_{00};\quad H_{BW}^{(1)}=\sum_{n\neq 0}\frac{H_{0n}H_{n0}}{n\omega};\quad H_{BW}^{(2)}=\sum_{n_1,n_2\neq 0}\left(\frac{H_{0,n_1}H_{n_1,n_2}H_{n_2,0}}{n_1n_2\omega^2}-\frac{H_{0,n_1}H_{n_1,0}H_{0,0}}{n_1^2\omega^2}\right)
\end{equation}
We see that the expressions have the advantage that the third-order term is more compact than in the van Vleck expansion. Their disadvantage compared to both the van Vleck expansion and the Magnus expansion is that the expressions are not in terms of nested commutators. Therefore, terms can quickly become non-local.

\subsubsection{Perturbative construction of a rotating frame}
\label{sec:perturb_rot_frame}

In this section we will review the approach introduced in Ref. \cite{PhysRevB.95.014112}, where the Floquet Hamiltonian $H_F$ is constructed using the relation $H_F=i\log(U(T))/T$, so we want to find $\log(U(T))$. The starting point is the  Schr\"odinger equation with the Hamiltonian $H(t)=H_0+V_0(t)$. After the application of a unitary transform $Q(t)$ it becomes
\begin{equation}
i\partial_t \psi=H_{\mathrm{eff}}\psi;\quad H_{eff}(t)=Q^\dag(t)( H_0+V_0(t)-i\partial_t)Q(t),
\end{equation}
where we find an effective Hamiltonian $H_{eff}(t)$. We want to choose the unitary transformation such that the effective Hamiltonian is time independent, that is go to an appropriate rotating frame. This can be achieved perturbatively by the following  steps.

We choose $Q(t)=e^{\hat \Omega(t)}$. Then making use of the derivative of the exponential map the effective Hamiltonian may be written as
\begin{equation}
    H_{\mathrm{eff}}= G(t)-i\partial_t\hat \Omega;\quad G(t)=e^{ad_{\hat \Omega}} H(t)-i\frac{1-e^{-ad_{\hat \Omega}}}{ad_{\hat \Omega}}\partial_t\Omega+i\partial_t\hat \Omega,
\end{equation}
where $ad_{\hat \Omega}=[\hat\Omega,.]$ and we have defined $G(t)$. This expression still depends on a so-far undetermined $\hat \Omega$ that can be chosen freely. We may construct a $\hat \Omega$ that determines the Floquet Hamiltonian if we choose it such that it removes the time dependent part $\tilde V(t)=G-1/T\int_0^T dt G(t)$ of $G(t)$ by requiring
\begin{equation}
\tilde V(t)-i\partial_t\hat\Omega=0.
\label{removeVeq}
\end{equation}
This can be fulfilled perturbatively if we expand $\hat\Omega=\sum_{n=1}^\infty \hat\Omega_n$ and assume that $\hat\Omega_n=\mathcal{O}(\omega^{-n})$ as well as $\partial_t\hat \Omega_n=\mathcal{O}(\omega^{-n+1})$. Such an expansion then allows us to also choose $G=\sum_{n=0}^\infty G_n$ with $G_n=\mathcal{O}(\omega^{n-1})$. Collecting terms of the same order in the definition of $G(t)$ one explicitly finds \cite{PhysRevB.95.014112} 
\begin{equation}
    G_n=\sum_{k=1}^{n} \frac{(-1)^k}{k!} \sum_{\begin{subarray}{l}{1\leq i_1,...,i_k\leq n} \\ { i_1+...+i_k=n} \end{subarray}} 
{\rm ad}_{\hat\Omega_{i_1}}...{\rm ad}_{\hat \Omega_{i_k}} \, H(t) +
 i\sum _{m=1}^n \sum_{k=1}^{n+1-m} \frac{(-1)^{k+1}}{(k+1)!} \sum_{\begin{subarray}{l}{1\leq i_1,...,i_k\leq n+1-m} \\ { i_1+...+i_kn+1-m} \end{subarray}} 
{\rm ad}_{\hat\Omega_{i_1}}...{\rm ad}_{\hat\Omega_{i_k}} \, \partial_t \hat\Omega_m. 
\end{equation}

In this case one may also expand $\tilde V=\sum_{n=0}^\infty V_n$ and has $ V_n=G_n-1/T\int_0^T dt G_n$. This allows us to perturbatively reduce Eq.\eqref{removeVeq} to
\begin{equation}
V_n(t)-i\partial_t\hat\Omega_{n+1}=0.
\end{equation}

One may then solve this problem iteratively. First, one may determine $\Omega_n=i\int dt V_n(t)$, then from there $G_n$, and finally $V_n(t)$. The effective Hamiltonian ultimately is given as
\begin{equation}
    H_{\mathrm{eff}}=\sum_{n} \frac{1}{T}\int dt G_n.
\end{equation}

This expansion at least to low orders agrees with the Magnus expansion\cite{PhysRevB.95.014112} and shares its drawbacks. However, it is not clear if both expansions agree to all orders.

\subsection{Mid-frequency approximations}

One may now wonder why we do not work with the different high-frequency approximations and construct as many orders as needed since the general expressions are known. The reason for this is two-fold. First, nested commutators become increasingly cumbersome to compute. Second, the expansions for quite generic situations are asymptotic expansions and therefore have an optimal cut-off order $n^*$ that shrinks with shrinking frequency $n^*\propto \omega$ \cite{PhysRevB.95.014112}. Thus even to reach intermediate frequency regimes, one has to resort to other techniques. To be precise, an intermediate frequency regime is one where either the local driving strength $v\ll\omega$ or the local strength of the constant part of the Hamiltonian $h_0\ll\omega$ conditions are not satisfied. In the next subsections, we will review some of the standard techniques used in the literature to approach this problem.

\subsubsection{Replica approximation}
We start the section by reviewing an approximation that is valid for a particular type of Floquet system with a Hamiltonian of the form
\begin{equation}
    H(t)=\begin{cases}
    H_0 & nT<t<nT+t_0\\
    H_1& nT+t_0<t<(n+1)T
    \end{cases}.
\end{equation}
The Floquet Hamiltonian in this case is given (using $t_1=T-t_0$ as a shorthand) as
\begin{equation}
    H_F=i\log(U);\quad U=e^{-iH_0t_0}e^{-iH_1t_1}.\quad
\end{equation}
In the case that $t_0$ and $t_1$ are sufficiently small an approximate result can be found by the Magnus expansion or equivalently for this special discrete case by the Baker Campbell Haussdorf identity $\log(e^{X}e^{Y})=(X + Y  + \frac{1}{2}[X,Y]+...)$, which is an expansion in both small $t_0$ and $t_1$. 

For only $t_1$ being sufficiently small one may expect that there could be an expansion in orders of $t_1$ only. Formally such an expansion can always be found as
\begin{equation}
    H_F=i\sum_{n}\frac{t_1^n}{n!}[\partial_{t_1}^n \log(U)]_{t_1=0},
\end{equation}
which, however, in its current form because of the operator logarithm is too cumbersome to work with. Luckily this problem can be avoided to an extent by following the approach in Ref.\cite{PhysRevLett.120.200607}, where the replica trick
\begin{equation}
  \log (U)=\lim_{\rho\to 0}\frac{1}{\rho}(U^\rho -1),
\end{equation}
was used to compute derivatives $[\partial_{t_1}^n \log(U)]_{t_1=0}$ as $[\partial_{t_1}^0 \log(U)]_{t_1=0}=(-i)t_0H_0$ and
\begin{equation}
    [\partial_{t_1}^n \log(U)]_{t_1=0}=(-i)^n\lim_{\rho\to 0}\frac{1}{\rho}\left[\sum_{0\leq m_1<...m_n<\rho}c_{m_1...m_r}\left(\prod_{j=r}^1 H_1(m_jt_0)\right) \right],
    \label{part_deriv_logU}
\end{equation}
where the assumption was made that derivatives commute with the limit and that under the limit $U^\rho\to \mathbb{1}$. The shorthand notations $H_1(t)=e^{it \; ad_{H_0}} H_1$ for the interaction picture Hamiltonian and $c_{m_1...m_r}=\frac{r!}{n_0!n_1!\cdots n_r!}$ for the multi-nomial coefficient, where $n_q$ is the number of indices $m_j=q$, were used.

The sum in Eq.\eqref{part_deriv_logU} has to be computed as a formal summation up to an arbitrary value of $\rho$ and then the limit is taken by assuming an analytic continuation to values of $\rho\to 0$.  This approximation can be valid as long as $t_1$ is sufficiently small. It is a medium frequency approach because it can reach periods $T$ longer than those found in the high-frequency approach. However, even for small $t_1$ it does not work for arbitrarily large $T=t_1+t_2$, which can be seen for example in Ref. \cite{vogl2019}.  

The advantage of the expansion is that it has a clean expansion parameter despite being valid in the medium frequency regime, a property that other expansions that we will discuss in the following do not share. This feature made it possible to find an optimal cut-off for the series associated with an estimate of the length for a pre-thermal regime \cite{PhysRevLett.120.200607}. Its disadvantage is that the approximation is quite cumbersome to compute and is limited to specific shapes of periodic drives.

\subsubsection{Non-perturbative rotating frame approaches}
\label{sec:nonperturbRotFrame}

We next review an approach to the mid- frequency regime that is inspired by our intuition about rotating systems --observing a rotating system in a co-moving frame makes it appear static. Again, this reduces to the question that we posed in section \ref{sec:perturb_rot_frame}: How to choose a unitary transformation $Q(t)$ such that a transformed Hamiltonian $H_{\mathrm{eff}}(t)=Q^\dag(t)(H(t)-i\partial_t)Q(t)$ is closer to the Floquet Hamiltonian? That is, how to appropriately choose a rotating frame to aid in the construction of a approximate Floquet Hamiltonian $H_{\mathrm{eff},0}=\frac{1}{T}\int_0^T dt H_{\mathrm{eff}}(t)$? This question was studied in various publications \cite{bukov2015,vogl2019,vogl2020effective,Vogl_2019AnalogHJ,rodriguezvega_2020a,PhysRevB.100.104306}. 

It is useful to first find constraints that can be employed to make educated guesses. In our case, there are two helpful constraints. First, we recognize that one might want to choose $Q(t)$ such that the time-dependent part of $H_{\mathrm{eff}}(t)=H_{\mathrm{eff},0}+V_{\mathrm{eff}}(t)$ is smaller than in the case of $H(t)=H_0+V(t)$,
\begin{equation}
\lVert V_{\mathrm{eff}}(t)\rVert<\lVert V(t)\rVert,
\end{equation}
where $\lVert.\rVert$ is an appropriately chosen operator norm--for instance the Frobenius norm. In such a case one finds the approximation is better if one replaces $H_{\mathrm{eff}}(t)\to \frac{1}{T}\int_0^T dt H_{\mathrm{eff}}(t)$ than in the case of $H(t)\to \frac{1}{T}\int_0^T dt H(t)$.

Secondly, one recognizes that the time-evolution operator $U_{\mathrm{old}}(t)=\mathcal{T}e^{-i\int_0^t H(t)}$ in the original frame is related to the time evolution operator in the new frame $U_{\mathrm{new}}(t)=\mathcal{T}e^{-i\int_0^t H_{\mathrm{eff}}(t)}$ by
\begin{equation}
    U_{\mathrm{old}}(t)=Q(t)U_{\mathrm{new}}(t).
\end{equation}
This relation tells us that if $Q(T)=\mathbb{1}$ then we have
\begin{equation}
    H_F=i\log(U_{\mathrm{old}}(T))/T=i\log(U_{\mathrm{new}}(T))/T,
\end{equation}
and therefore we should require 
\begin{equation}
Q(T)=\mathbb{1}.
\end{equation}

The above two restrictions for $Q(t)$ suggest the unitary transformation
\begin{equation}
    Q(t)=e^{-i\int_0^t dt V(t)},
\end{equation}
which exactly removes $V(t)$ if it is sufficiently small or if $[V(t),V(t^\prime)]=0$ and $\lVert H_0\rVert/\lVert V(t)\rVert\to 0$. However, even outside those two limits this approach often offers an improvement over the high frequency expansions \cite{bukov2015,vogl2019}.  An approximate Floquet Hamiltonian is then given as $H_F\approx \frac{1}{T}\int_0^T dt H_{\mathrm{eff}}(t)$.

The unitary transformation above is not the only possibility, but many choices are available and are associated with varying degrees of success. For instance, one can also decompose $H(t)=H_0+\sum_nV_n(t)$ with freedom on how to choose the $V_n(t)$. One can then construct a unitary transformation
\begin{equation}
Q(t)=\prod_n e^{-i\int_0^t dt V_n(t)}.
\end{equation}

In some cases certain choices on how to decompose $V(t)$ into $V_n(t)$ can be more advantageous than others: A specific choice may allow keeping symmetries that are not retained by other choices when the average $H_F\approx \frac{1}{T}\int_0^T dt H_{\mathrm{eff}}(t)$ is taken. In Ref. \cite{vogl2020effective,rodriguezvega_2020a} the example of a periodically driven twisted bilayer graphene and twisted double bilayer graphene are discussed, where this approach makes it possible to derive a non-perturbative Floquet Hamiltonian that keeps rotational symmetry in momentum space that would otherwise be broken. The advantage of the approach discussed in this section is the freedom to choose and that it has a clear physical interpretation. The disadvantage is that often the transformations can be challenging to compute.

\subsection{Low-frequency approximations}

As we have seen in the previous section, as the frequency of the drive decreases, it becomes increasingly difficult to find reliable approximation schemes for the Floquet Hamiltonian. This is especially true in the low frequency regime where local interactions satisfy $h \gg \omega$, where multiple Floquet zones intersect each other. Here we present three approaches: a self-consistent low frequency approach, a flow equation approach, and a Floquet perturbation theory.

\subsubsection{Floquet perturbation theory}
\label{sec:Floquetperturbationtheory}
In this section, we review the Floquet perturbation theory~\cite{Rodriguez_Vega_2018}. This theory is derived for periodically driven systems, but it can be applied to more general drives by appropriately choosing the shape of the drive over a single cycle. In this sense, this approach can be compared with adiabatic perturbation theory~\cite{teufel2003adiabatic, PhysRevA.78.052508, PhysRevLett.104.170406}, and adiabatic-impulse theory~\cite{SHEVCHENKO20101, PhysRevB.97.205415}. 

First, we will introduce useful notation to describe the extended space, and then we will derive the perturbation expansions. This approach's main advantage is that it allows us to describe both the high- and low-frequency regimes on equal footing. 

\textit{Definitions}

In Sec. \ref{sec:From the perspective of a quasi-energy operator}, we showed that a time-dependent Floquet problem can be cast into a time-independent problem in an extended Hilbert space $F=H\otimes I$, where H is the Hilbert space of the original time-dependent system and $I$ is an auxiliary space spanned by a complete set of bounded periodic functions over $[0,T)$~\cite{PhysRevA.7.2203}. A choice of such basis functions that is particularly useful for our purposes is $\{|t)\}, 0\leq t<T=2 \pi/\Omega$, which satisfy the orthogonality relation $(t'| t) = T\delta( t-t')$. The periodicity of the functions $|t)$ allow us to expand in a Fourier basis $|n) = \int_0^T e^{-i n\Omega t}|t)\frac{dt}{T}$,  $n\in\mathbb{Z}$ with $(n|m) = \delta_{nm}$. The basis $\{ |n)\}$ is the same basis we discussed in Sec. \ref{sec:From the perspective of a quasi-energy operator}, but here it is more convenient to work with its Fourier transform $|t)$. 

States in $F$ are constructed as $\kket{\phi_t}:=\ket{\phi(t)}|t)$, were $\ket{\phi(t)}$ are the periodic steady states. In $F$, the Floquet Schr\"odinger equation takes the form 
\begin{eqnarray}\label{eq:FSch}
(\hhat{H} - \hhat{Z}_{\mnote{t}})\kket{\phi_{\alpha n}} = \epsilon_{\alpha n} \kket{\phi_{\alpha n}},
\end{eqnarray}
where the Hamiltonian is given by $\hhat H = \int_0^T \hat H(t) \otimes |t)(t| \frac{dt}{T}$, $\hhat Z_{\mnote{t}}= \hat I\otimes\sum_{n\in\mathbb{Z}} |n) n\Omega (n|$. The Floquet states are constructed as $\kket{\phi_{\alpha n}} \equiv \hhat \mu_n\kket{\overline{\phi_\alpha}}$, with $\hhat \mu_n = \hat I\otimes\int_0^T |t)e^{{+}in\Omega t}(t|\frac{d t}{T}$ the ladder operator, and $\kket{\overline\phi} \equiv \int_0^T \kket{\phi_t}\frac{dt}{T}$. Finally,  $\epsilon_{\alpha n} \equiv \epsilon_\alpha{+}n\Omega$ are the quasienergies. Note that the ladder operator satisfies the commutation relations $[\hhat H,\hhat\mu_n] = 0$, and ${[ \hhat\mu_n, \hhat Z_{\mnote{t}}]} = n\Omega\hhat \mu_n$.

\textit{Low-Frequency Expansion}

For the construction of the low-frequency perturbation theory, we first re-scale the time $\tau = \Omega t$ and note that the adiabatic limit is obtained by dropping the term proportional to $\Omega$. We find
\begin{eqnarray}\label{eq:FH0}
\hhat H \kket{\psi_{\alpha \tau}} = E_{\alpha \tau} \kket{\psi_{\alpha \tau}},
\end{eqnarray}
where $\kket{\psi_{\alpha \tau}}=\ket{\psi_\alpha(\tau)}|\tau)$, $\hat H(\tau)\ket{\psi_{\alpha}(\tau)} = E_{\alpha}(\tau)\ket{\psi_\alpha(\tau)}$, and $E_{\alpha \tau}=E_{\alpha}(\tau)$. We assume that the static energy spectrum is not degenerate. Also, notice that the eigenstates are defined up to a phase, which is fixed by the definition of the zero-th order Floquet-Schr\"odinger equation~\cite{Martiskainen2015}. Explicitly, $
\ket{\kappa_\alpha(\tau)} = e^{-i\Lambda_\alpha(\tau)} \ket{\psi_\alpha(\tau)}$ with $\Lambda_\alpha(2\pi)=\Lambda_\alpha(0)$ is another basis, which after being plugged into the Floquet-Schr\"odinger equation leads to 
$ 
[\hat H(\tau) - i\Omega\frac{\partial}{\partial\tau}]\ket{\kappa_\alpha(\tau)}
- i\Omega e^{-i\Lambda_\alpha(\tau)}\frac{\partial}{\partial\tau}\ket{\psi_\alpha(\tau)}.
$ 
Thus, setting 
$ 
\Lambda_\alpha(\tau) =\frac1\Omega\int_0^\tau [E_\alpha(s)- \epsilon_{\alpha(0)}] ds/(2 \pi)$,
we find the zeroth-order Floquet Schr\"odinger equation
\begin{eqnarray}\label{eq:zeroF}
[\hat H(\tau) - i\Omega\frac{\partial}{\partial\tau}]\ket{\kappa_\alpha(\tau)}
	\approx \epsilon_{\alpha(0)}\ket{\kappa_\alpha(\tau)} .
\end{eqnarray}
Thus, $\epsilon_{\alpha(0)}$ is interpreted as the zero-th order approximation to the quasienergies. In the following expressions, we re-define $\ket{\psi_\alpha(\tau)}$ such that they include the appropriate gauge.

Since the existence of Floquet copies of quasienergies is a striking property of the structure of the extended space Hamiltonian that has been confirmed experimentally\cite{Mahmood2016}, it is important to preserve this property in an approximation, that is one has to require that eigenvalues are modular: $\epsilon_{\alpha} = \epsilon_{\alpha} + n \Omega $.
This is achieved by applying the ladder operator to $\kket{\psi_{\alpha\tau(i)}}$, the solution of order $i$, which can be found by applying conventional perturbation theory 
\begin{eqnarray}
\kket{\phi_{\alpha n(i)}} = \hhat\mu_n\kket{\overline{\psi_{\alpha(i)}}} = \int_0^{2\pi} e^{{+}in\tau}\kket{\psi_{\alpha \tau(i)}} \frac{d \tau}{2 \pi},
\end{eqnarray}
with modular eigenvalue $\epsilon_{\alpha n (i)} =  \epsilon_{\alpha (i)} {+} n\Omega$, and the quasienergy
\begin{eqnarray}
\epsilon_{\alpha (i)} = \bbra{\overline{\phi_{\alpha (0)}}}\hhat H\kket{\overline{\phi_{\alpha(i)}}} - \Omega \bbra{\overline{\phi_{\alpha (0)}}}\hhat Z\kket{\overline{\phi_{\alpha(i-1)}}}.
\label{eq:qe_fpt}
\end{eqnarray}

The expression for the quasienergies follows from the Floquet-Schr\"odinger equation and requiring normalization of the wavefunctions. By replacing the definitions of the steady states and the time-derivatives in the extended Hilbert space F in Eq. \eqref{eq:qe_fpt}, we find the expressions for the zero-th and first order corrections to the quasienergies
%

\begin{align}
\epsilon_{\alpha(0)} 
	&= \int_0^{2\pi} E_{\alpha}(\tau) \frac{d\tau}{2 \pi}, \label{eq:FAPTe0}\\
\epsilon_{\alpha(1)}
	&=\Omega\int_0^{2\pi} \bra{\psi_\alpha(\tau)}\frac1i\frac{\partial}{\partial\tau}\ket{\psi_\alpha(\tau)} \frac{d\tau}{2 \pi} \label{eq:FAPTe1}. 
\end{align}
Likewise, the first order correction to the steady states is given by
\begin{align}
\kket{\phi_{\alpha n(1)}} 
	& = \Omega \int_0^{2\pi} \sum_{\beta\neq \alpha} \frac{\bra{\psi_\beta} \frac1i\frac{\partial}{\partial \tau}\ket{\psi_\alpha}}{E_\alpha(\tau)-E_\beta(\tau)}e^{{+}i n\tau}\kket{\psi_{\beta\tau}} \frac{d\tau}{2 \pi}. \label{eq:FAPTp1}
\end{align}

Therefore, by introducing appropriate definitions, we can construct a Floquet perturbation theory in the low-frequency regime by systematically introducing corrections to the adiabatic limit. In the next section, we address the degenerate case. \\

\begin{figure}[]
	\begin{center}
		\includegraphics[width=8.0cm]{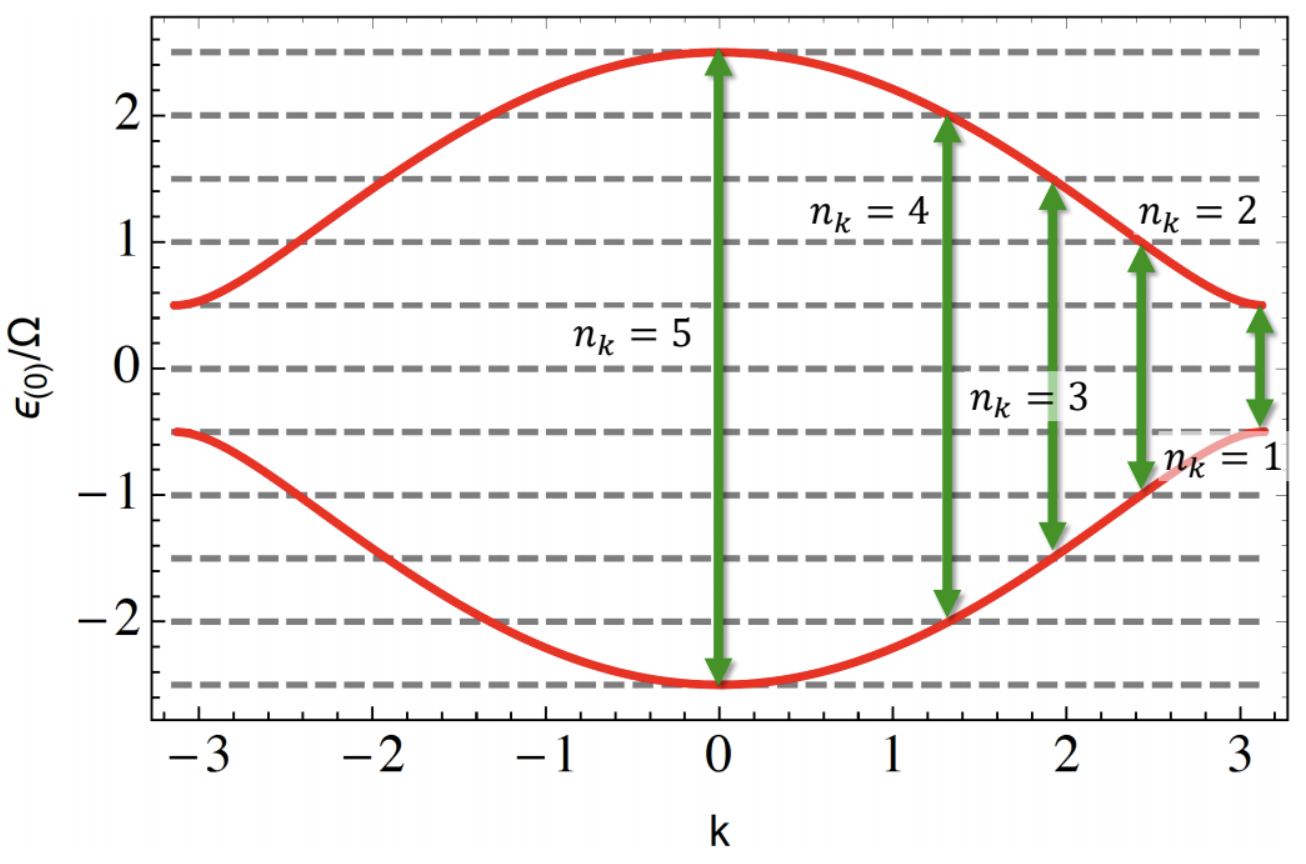}
		\caption{(Color online) Zeroth-order quasienergy as a function of momentum $k$ for a model system that illustrate the degeneracies arising in the low-frequency regime.}
	\end{center}
	\label{fig:fpt_fig1}
\end{figure}

\textit{Degenerate Low-Frequency Floquet Perturbation Theory}\\
We have reviewed how quasienergies in the low frequency regime are constructed by introducing corrections to the adiabatic limit when dealing with non-degenerate systems. For a generic Floquet system, however, often there can be degeneracies--for example when bands of different Floquet copies intersect as illustrated in Fig. \ref{fig:fpt_fig1}. Therefore, it is important to also study what happens in such a case, which we will review in this section following \cite{Rodriguez_Vega_2018}. When there are degeneracies present in the spectrum, we need to employ degenerate Floquet perturbation theory to describe the drive's effects correctly. This concept has been used to describe heating processes in optical lattices\cite{Str_ter_2016,PhysRevA.92.043621}. 

Degeneracies can be present in the static energy spectrum, or arise from the overlap of Floquet copies of the spectrum. For example, consider a two-band system with \emph{non-degenerate} instantaneous energy eigenvalues $E_\alpha \neq E_\beta$. In Fig. \ref{fig:fpt_fig1}, we plot the unfolded zeroth order quasienergies, Eq. \eqref{eq:FAPTe0}, as a function of momentum. Degeneracies arise when $\epsilon_\alpha = \epsilon_\beta + n_k\Omega = m \Omega/2$ for integers $n_k,m$. For accidental or symmetry related degeneracies, $n_k=0$, one can restrict the unperturbed quasienergies to the first Floquet zone. 

Define a state 
$
\kket{\chi_{\alpha}^r} = \sum_s c_\alpha^{rs} \hhat\mu_{n^r}\kket{\overline{\phi_{\alpha(0)}^s}},
$
with coefficients $c_\alpha^{rs}$ to be determined by solving
\begin{eqnarray}
(\hhat H - \Omega \hhat Z) \kket{\chi_{\alpha}^r} = (m\Omega/2 + \epsilon_{\alpha(1)}^r)\kket{\chi_{\alpha}^r} + O(\Omega^2).
\end{eqnarray}
This yields,
\begin{eqnarray}\label{eq:dFPTA1}
\sum_q W_\alpha^{rq} c_\alpha^{qs} = \epsilon_{\alpha(1)}^{s} c_\alpha^{rs},
\end{eqnarray}
where $W_\alpha$ is a matrix with diagonal elements $W_\alpha^{rr} = \epsilon_\alpha^r {+} n^r\Omega - m\Omega/2$, and the off-diagonal elements,
\begin{eqnarray}
W_\alpha^{r s} 
	&\equiv -\Omega \bbra{\overline{\phi^r_{\alpha}}} \hhat\mu_{n^r}^\dagger \hhat Z \hhat\mu_{n^s}\kket{\overline{\phi^s_{\alpha}}}
	= \Omega \int_0^{2\pi} e^{i({n_s-n_r})\tau} \bra{\phi_\alpha^r(\tau)}\frac1i\frac{\partial}{\partial\tau}\ket{\phi_\alpha^s(\tau)}  \frac{d\tau}{2 \pi}, \quad r\neq s.
\end{eqnarray}
The eigenvalues of the equations in Eq.(\ref{eq:dFPTA1}) are the first-order degenerate low-frequency Floquet perturbation theory. 

This approach is simple to implement, and higher-order corrections can readily be obtained following the usual perturbation theory taking advantage of appropriate definitions. It is most useful when we are interested in the quasienergies and steady states directly, since it cannot be used to construct a closed-form effective Hamiltonian. In the next sections, we will discuss how such effective Floquet Hamiltonian valid in the low-frequency regime can be obtain.

\subsubsection{Self consistent low frequency approach}
\label{sec:selfconsistentlow}

In Sec. \ref{sec:Floquetperturbationtheory}, we introduced a theory to construct the quasienergies and steady-states perturbatively, by adding corrections to the adiabatic limit. In this section, we introduce a non-perturbative approach which allows us to derive effective Floquet Hamiltonians in the weak-drive limit, valid for arbitrary frequency \cite{Vogl2020_effham}. \\

\textit{Time analog of the empty lattice picture}. \\

\begin{figure}[]
	\begin{center}
		\includegraphics[width=10.0cm]{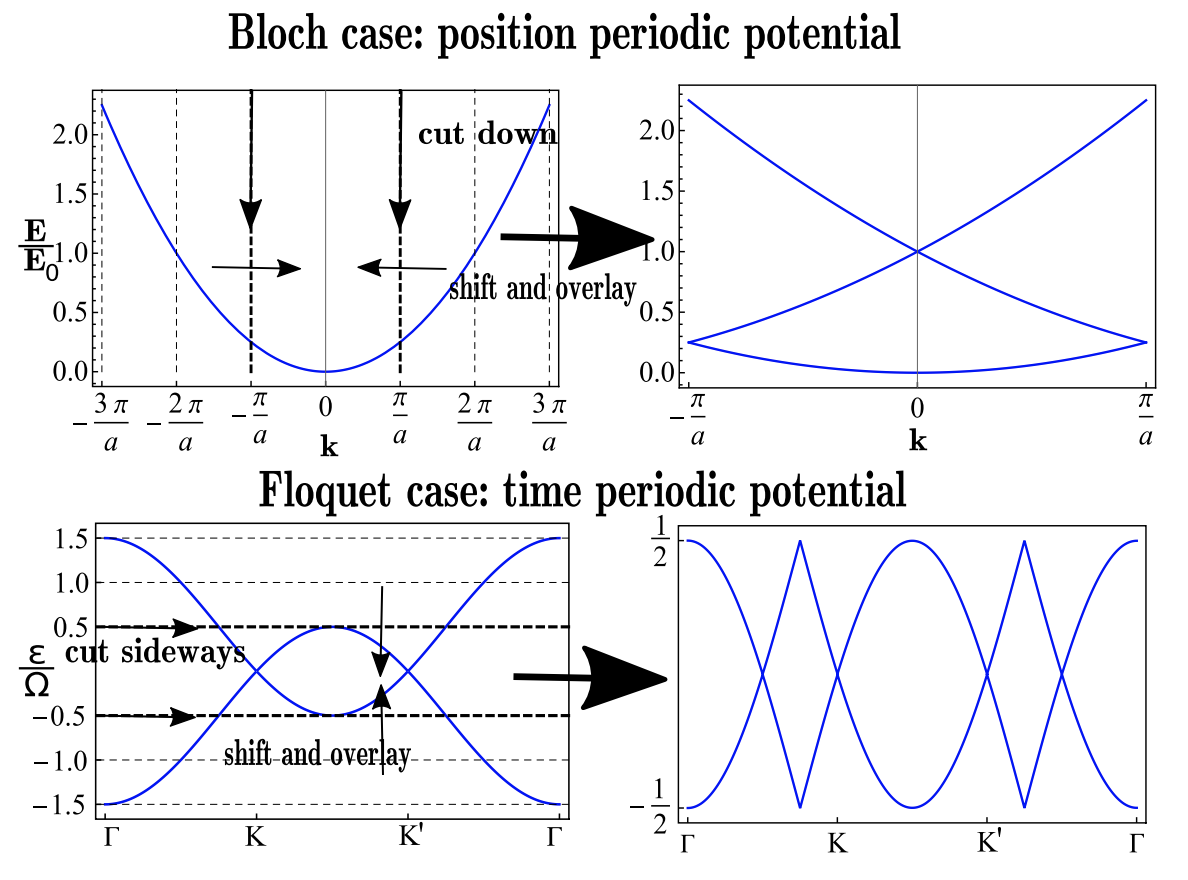}
		\caption{(Color online) Zone folding scheme for the empty lattice picture. For spatially-periodic potentials the momentum axis is folded in, while for time-periodic potentials the energy axis is folded. Reprinted from Ref. \cite{Vogl2020_effham}.}
	\end{center}
	\label{fig:low-freq_fig1cartoon}
\end{figure}

To understand the effect of a weak periodic drive, it is useful to make an analogy with the effect of weak spatially-periodic potentials. Consider the Hamiltonian of a free electron in one spatial dimension $H_0$. In the presence of an infinitesimally-weak periodic potential $V(x) = V(x+a)$, the usual parabolic energy-momentum relation is \textit{folded} into the first Brillouin zone (BZ) of size $2\pi/a$ leading to a complicated collection of energy bands~\cite{kittel2004introduction,ashcroft1976solid,marder2010condensed}. Fig. \ref{fig:low-freq_fig1cartoon} shows schematically the effect of $V$. In some materials, this crude approximation can lead to good approximations~\cite{PhysRev.118.1182}.  
The same idea can be used for time-periodic Hamiltonians, $H(t) = H(t + 2\pi/\Omega)$. The periodicity of the drive induces a set of copies of the energies that can overlap for small $\Omega$, compared with the bandwidth of the system $W$. Fig. \ref{fig:low-freq_fig1cartoon} demonstrates schematically the effect. Notice that the band-crossings can become avoided crossings depending on the details of the system.   

We consider a monochromatic drive in the weak-drive limit, characterized by drive amplitude $A \ll 1$. The Hamiltonian takes the general form
\begin{equation}
 h(t)=  h_0+ P(A) e^{-i\Omega t}+ P^\dag(A) e^{i\Omega t},
\label{eq:monocromatic}
\end{equation}
where $h_0$ is the static Hamiltonian, and $P(A)$ the first-harmonic operator, which is a function of the drive amplitude $A$. As we discussed generally in Sec. \ref{sec:From the perspective of a quasi-energy operator}, in the extended space, the Floquet-Schr\"odinger equation can be written as~\cite{2015PhRvA..91c3416P,10.1088/2515-7639/ab387b,PhysRevA.7.2203,Eckardt_2015,PhysRevA.95.023615,PhysRevLett.111.175301,Rodriguez_Vega_2018} 
 \begin{align}
 	&\begin{pmatrix}
 	\ddots&\vdots&\vdots&\vdots&\vdots&\vdots&\\
 	\cdots&P^\dag& h_0-\Omega&P&0&0&\cdots\\
 	\cdots&0&P^\dag&  h_0&P&0&\cdots\\
 	\cdots&0&0&P^\dag& h_0+\Omega&P&\cdots\\
 	&\vdots&\vdots&\vdots&\vdots&\vdots& \ddots
 	\end{pmatrix}\begin{pmatrix}
 	\vdots\\
 	\phi_{-1}\\
 	\phi_0\\
 	\phi_1\\
 	\vdots
 	\end{pmatrix}=\epsilon \begin{pmatrix}
 	\vdots\\
 	\phi_{-1}\\
 	\phi_0\\
 	\phi_1\\
 	\vdots
 	\end{pmatrix}.
 	\label{quasi-en-eq2}
 \end{align}
This equation can be decoupled \cite{2015PhRvA..91c3416P,10.1088/2515-7639/ab387b} into an equation for the first Floquet mode $\phi_0$ only. The result is the continued fraction
 \begin{equation}
 \begin{aligned}
h_{\mathrm{eff}}(\epsilon)=h_0+P\frac{1}{\epsilon-h_0-\Omega-P\frac{1}{\epsilon-h_0-2\Omega-\cdots}P^\dag}P^\dag
+P^\dag\frac{1}{\epsilon-h_0+\Omega-P^\dag\frac{1}{\epsilon-h_0+2\Omega-\cdots} P} P.
 \end{aligned}
 \end{equation}
Truncating to linear order in $P$ we obtain
\begin{equation}
 	h_{\mathrm{eff}}(\epsilon)\approx h_0+P\frac{1}{\epsilon-h_0-\Omega }P^\dag+P^\dag\frac{1}{\epsilon-h_0+\Omega} P,
 	\label{truncfrac}
\end{equation}
which defined the effective Floquet Hamiltonian in the weak-drive regime. Notice that there is no restriction on the frequency $\Omega$. To determine the quasi-energy spectrum and steady-state mode $\phi_0$, we need to solve the self-consistent Floquet-Schr\"odinger eigenvalue equation
\begin{equation}
    \left( h_{\mathrm{eff}}(\epsilon) - \epsilon \right) \phi_0=0.
\end{equation}
The rest of the Floquet modes $\phi_n$ for $|n| > 0$ are constructed using the recursion relation \cite{2015PhRvA..91c3416P,10.1088/2515-7639/ab387b}
\begin{equation}
(\epsilon+m\Omega-h_0)\phi_m=P^\dag \phi_{m-1}+P\phi_{m+1}.
\label{recursion}
\end{equation}
To leading order in $P$, be approximated as $ \phi_{\pm n}=\left(\epsilon \pm n\Omega-h_0\right)^{-1} P^\dag \phi_{\pm n \mp 1}$.

This method provides us with a finite representation of the Floquet problem, formally defined in an infinite representation in the extended space. The projection procedure results in a time-independent self-consistent Sch\"odinger-type equation for the quasienergies and the steady states. Furthermore, the effective Hamiltonian can be used to address other complex effects such as disorder which requires taking several averages. However, since we require to invert a matrix analytically to derive a closed-form expression for the effective Hamiltonian, this procedure works best for time-dependent problems initially defined in small Hilbert spaces, such as driven graphene.

\subsubsection{Real time flow equation approach}
In section \ref{sec:nonperturbRotFrame} and \ref{sec:perturb_rot_frame} we discussed how unitary transformations $Q(t)$ can be used to reduce the time dependent part $V(t)$ of a time periodic Hamiltonian $H(t)=H_0+V(t)$ to find a better approximation to the Floquet Hamiltonian. In section \ref{sec:perturb_rot_frame} we reviewed a perturbative high frequency approach to construct such a transformation. In section \ref{sec:nonperturbRotFrame} we reviewed general properties such a transformation needs to fulfill--namely that (i) $V(t)$ shrinks by applying the unitary transform and that (ii) $Q(T)=\mathbb{1}$--to guess various transformations. In this section we will review an approach that allows us to construct such a unitary transform in terms of infinitesimal steps. We will follow the approach taken in Refs.\cite{vogl2019,Vogl_2019AnalogHJ}. 

As our starting point we recall that applying a time-dependent unitary transformation $Q(t)$ to the Schr\"odinger equation $i\partial_t\psi =H(t)\psi$ again results in the same equation just with the effective Hamiltonian
\begin{equation}
    H_{\mathrm{eff}}=Q^\dag(t)(H(t)-i\partial_t)Q(t).
\end{equation}
Because we want to transform the Hamiltonian in infinitesimal steps we restrict ourselves to an infinitesimal unitary transformation $Q(t)=e^{i\delta s\Omega(t)}$, where $\delta s$ is infinitesimal and $\Omega(t)$ hermitian to keep unitarity. To lowest order in a Taylor series we then find
\begin{equation}
    H_{\mathrm{eff}}=H(t)+\delta s\partial_t \Omega(t)+i\delta s[H(t),\Omega(t)].
    \label{inf_unitary_transf}
\end{equation}

One could now imagine that instead of one single infinitesimal unitary transformation one could also look at a full family of such transformations that are applied consecutively and determined by the operator $\Omega(t,s)$, where the parameter $s$ labels the different transformations. After each such transformation the Hamiltonian also changes and we can keep track of this by considering a family of Hamiltonians $H(t,s)$ that is also labeled by the parameter $s$. To fix the parameter $s$ by convention we set $H(t,s)=H(t)$--the Hamiltonian at $s=0$ is the original untransformed Hamiltonian. Using this notation we can then rewrite Eq. \eqref{inf_unitary_transf} as
\begin{equation}
    H(t,s+\delta s)=H(t,s)+\delta s\partial_t \Omega(t)+i\delta s[H(t,s),\Omega(t,s)].
\end{equation}
Expanding the left side to linear order in $\delta s$ we find a differential equation
\begin{equation}
    \frac{dH(t,s)}{ds}=\partial_t \Omega(t)+i[H(t,s),\Omega(t,s)].
    \label{general_flow-equ}
\end{equation}
Now we know how to chain together multiple consecutive infinitesimal unitary transformations. But our original goal was to construct a transformation that allows us to reduce the time dependent part of the Hamiltonian. In the following we will choose $\Omega(t,s)$ appropriately to do that.

If at every step we are able to split $H(t,s)=H_0(s)+V(t,s)$, then we can look to section \ref{sec:nonperturbRotFrame} for inspiration on how to construct a unitary transformation that fulfills requirements (i) and (ii) that we gave at the start of this section, namely one can choose  $\Omega(t,s)=-\int dt V(t,s)$, which will receive further justification in the following. With this we find that Eq.\eqref{general_flow-equ} can be simplified as
\begin{equation}
    \frac{dH(t,s)}{ds}=-V(t,s)+i\int_0^t dt^\prime [V(t^\prime,s),H(t,s)].
\end{equation}

Let us justify this choice. We find that for each infinitesimal unitary transform $Q(t,s)=e^{-i\delta s\int dt V(t)}$  $Q(T,s)=\mathbb{1}$ is fulfilled and therefore requirement (ii) is fulfilled. Additionally, at each step $V(t,s)$ is infinitesimally reduced via the first term $-V(t,s)$ and therefore requirement (i) is also fulfilled. Furthermore we find that the equation has a fixed point $\left.\frac{dH(t,s)}{ds}\right|_{s=s^*}=0$ once a point $s^*$ is reached for which $V(t,s^*)=0$. Therefore, the equation will for the most generic case flow to a point of $V(t,s^*)=0$.

This operator equation can be solved by means of solving a set of first order differential equations (flow equations) for coefficients $\{c^{(0)}_n,c^{(v)}_n\}$ if one chooses an ansatz as an operator sum for the Hamiltonian $H(t,s)= \sum c^{(0)}_n(s) \hat O^{(0)}_n+c^{(v)}_n(s) \hat O^{(v)}_n(t)$. In this $\hat O^{(0)}_n$ are time independent operators and $\hat O^{(v)}_n(t)$ are periodically time dependent operators. For instance, for a two level system one could have operators $\hat O^{(0)}_n(t)\in \{\sigma_x,\sigma_y,\sigma_z\}$ and $\hat O^{(v)}_n(t)\in\{e^{in\omega t}\sigma_x,e^{in\omega t}\sigma_y,e^{in\omega t}\sigma_z\}$ with $n\in \mathbb{Z}\setminus\{0\}$. The operators are best chosen linearly independent. They can either be linearly independent because of the operator itself (as an example $\sigma_x$ and $\sigma_y$ are already linearly independent according to the Frobenius inner product) or because of the time-dependent function that is associated with the operator (as an example $\sigma_x$ and $\sigma_xe^{i\omega t}$ are linearly independent because $1$ and $e^{i\omega t}$ are orthogonal functions). It is important to note that often it might be necessary to truncate the ansatz of the form $H(t,s)= \sum c^{(0)}_n(s) \hat O^{(0)}_n+c^{(v)}_n(s) \hat O^{(v)}_n(t)$ for instance because infinitely terms would be generated. This can be done by physical intuition, perturbative insights, and symmetries.

\begin{figure}[t]
	\begin{center}
	\includegraphics[width=8.50cm]{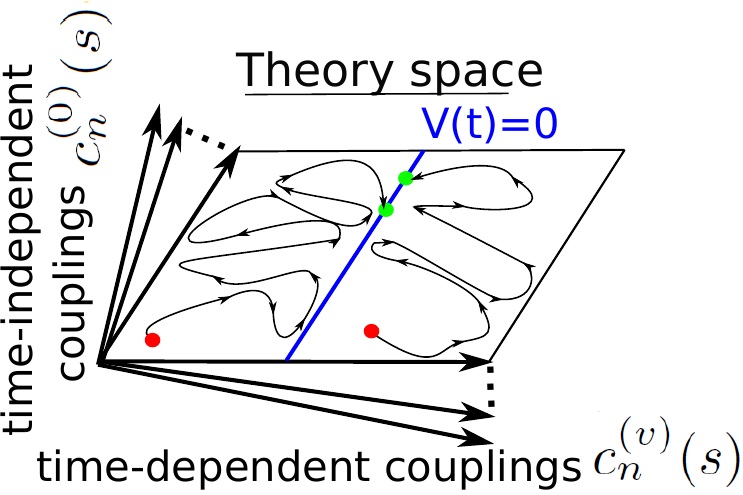}	
		\caption{(Color online) Figure that demonstrates how couplings for time dependent operators  $c^{(v)}_n$ and couplings for time-independent operators $c^{(0)}_n$ flow in theory space until a Hamiltonian on the fixed point line is reached where only couplings for time-independent couplings $c^{(0)}_n$ remain.}
		\label{coupling_flow}
	\end{center}
\end{figure}

To gain further intuition on how a Hamiltonian changes along the flow we have schematically depicted in figure \ref{coupling_flow} how its couplings may flow in theory space.
One finds that generically the coefficients for the Hamiltonian will flow in theory space until a theory is reached where $V(t,s)=0$, which is a "line" of fixed points. One should, however, point out that before such a fixed point is reached, especially for low frequency drives, different fixed points can be approached quite closely. In Ref.\cite{vogl2019} this is discussed in detail for the example of the Schwinger-Rabi model $H=B_z\sigma_z+B\cos(\omega t)\sigma_x+B\sin(\omega t)\sigma_y$.

The advantage of the approach we described in this section is that in principle it can be exact as long as there are no non-generic fixed points with $V(t,s^*)\neq 0$--and we are not aware of any such cases. The disadvantage is the analytical effort that is needed to set up flow equations for the coefficients.

\subsubsection{Extended space picture flow equation approach}

Instead of working directly with a time-dependent Hamiltonian one can also work in  the extended space picture description from section \ref{sec:From the perspective of a quasi-energy operator}. In this case one has the quasi-energy equation
\begin{equation}
    \epsilon\psi=\hat Q\psi;\quad \hat Q=\hat Q_D+\hat Q_X,
\end{equation}
where $\hat Q_D=\mathcal{H}_D+\omega\mathcal{M}$  with $[\mathcal{M}]_{mn}=m\delta_{mn}$ and $[\mathcal{H}_D]_{mn}=\delta_{mn}H_{m-n}$ describes the diagonal  (in photon space) part of the problem and $Q_X=$ with $[\mathcal{H}_X]_{mn}=(1-\delta_{mn})H_{m-n}$ the off-diagonal part. We recall that
$H_{n}=\frac{1}{T}\int_0^T dt e^{-in\omega t}H(t)$.  To get an effective Hamiltonian one may want to get rid of the off-diagonal component $\hat Q_X$ and this could be done by infinitesimal unitary transformations that we explain how to work with in the following. In this section we review the approach by Ref.\cite{verdeny2013}.

One may assume that $\hat Q$ belongs to a family of operators $\hat Q(s)$. Since this is a  time independent problem we may applying a parameter $s$-dependent infinitesimal unitary transformation as $\hat Q(s+\delta s)=e^{i\delta s\Omega(s)}\hat Q(s)e^{-i\delta s\Omega(s)}$. A Taylor expansion on both sides in $\delta s$ allows us to find
\begin{equation}
    \frac{d\hat Q(s)}{ds}=i[\Omega(s),\hat Q(s)],
\end{equation}
which tells us how the quasi energy operator $\hat Q(s)$ transforms under the application of a parameter $s$ dependent chain of infinitesimal unitary transformations $e^{-i\delta s\Omega(s)}$.

Now it is known from Ref.\cite{verdeny2013} that off-diagonal terms in such an operator $\hat Q$ can be reduced if a generator $\hat Q$ is chosen as
\begin{equation}
    \Omega(s)=i\omega[\mathcal{M},\hat Q_X(s)].
\end{equation}
It is easy to see that for generic situations (unless non-generic fixed points $Q_X(s)\neq 0$ are encountered) this will lead to a transformed block-diagonal $\hat Q(s)$. This is because if $\hat Q_X(s)=0$ then $\frac{d\hat Q(s)}{ds}=0$ and a fixed point that is generic for this choice of $\Omega(s)$ is reached. One may  find flow equations for couplings $\{c_n^D(s),c_n^X(s)\}$ if an ansatz $Q(s)=\sum_n c_n^D(s)\hat O^D_n+c_n^X(s)\hat O^X_n$ is made. Here $\hat O^D_n$ denotes operators that take values on the diagonal blocks and $\hat O^X_n$ denotes operators that take values on the off-diagonal blocks of $\hat Q$. A more detailed discussion of this approach can be found in \cite{verdeny2013}. The difficulty associated with it is to keep track of all the commutators for the operators in the extended space. The advantage over the approach in the previous section is that one does not need to keep track of any integrals.

This method completes our overview of the theoretical tools available to obtain effective Floquet Hamiltonians in the three main frequency regimes. In the the next section, we employ some of these techniques to study light-driven quantum materials.

\section{Applications to quantum materials}

In the previous sections, we reviewed the theoretical state-of-the-art tools available to describe periodically-driven systems, which is the main objective of this work. In this section, we review recent applications of these techniques to two types of quantum systems: moir\'e superlattices and strongly-correlated systems driven with light pulses. There are several theoretical studies and proposals for equilibrium Moir\'e heterostructures~\cite{Kennes2020,doi:10.1021/acs.nanolett.9b00986,xian2020realization,PhysRevResearch.2.022041,Wu_2018,Wu_2020,Hsu_2020,PhysRevResearch.2.033271,Pan_2020,PhysRevResearch.2.022010,PhysRevB.101.245436,PhysRevResearch.2.022010,PhysRevB.101.155149,moralesduran2020metalinsulator,shi2020moire,zhu2020curious,PhysRevB.102.115127,PhysRevLett.124.187601,PhysRevB.100.125104}. For comprehensive reviews on Moir\'e heterostructures in equilibrium see Refs.~\cite{kennes2020moire,andrei2020graphene}.

\subsection{Twisted bilayer graphene}

As a first example, we will review the previous work done for twisted bilayer graphene under the influence of light. 

\subsubsection{Static model for twisted bilayer graphene}

\begin{figure}[t]
	\begin{center}
		\subfigure{\includegraphics[width=8.50cm]{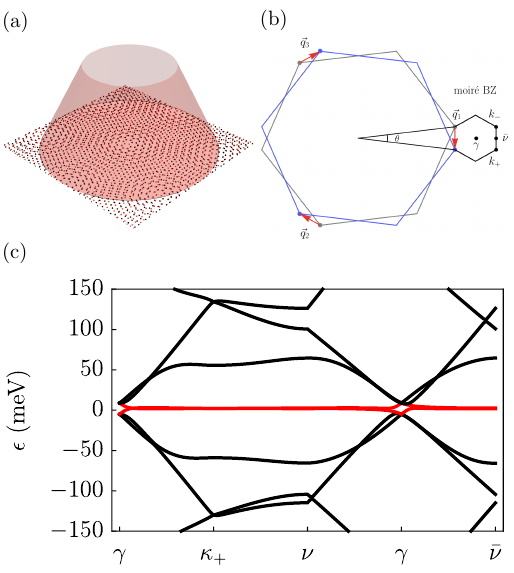}}		
		\caption{(Color online) (a) Sketch of twisted bilayer graphene irradiated by circularly polarized light. (b) moir\'e Brilloiun zone. (c) Band structure for twisted bilayer graphene for $w_0=w_1 = 110 $~meV, and $\theta=1.05^\circ$. The low-energy flat bands are highlighted in red. Figure reprinted from Ref. \cite{vogl2020effective}}
		\label{fig:twisted_fig1}
	\end{center}
\end{figure}

The low-energy physics of twisted bilayer graphene can be modeled with the continuum model derived by Bistritzer $\&$ MacDonald \cite{Bistritzer12233,Wu_2018,Rost_2019,fleischmann2019perfect,fleischmann2019moir,xie2018nature},
\begin{equation}
\begin{aligned}
&H_{\vect k}(\vect x)=\begin{pmatrix}
h(-\theta/2,\vect k-\kappa_-)&T(\vect x)\\
T^\dag(\vect x)&h(\theta/2,\vect k-\kappa_+)
\end{pmatrix}
\end{aligned}.
\label{twist_bilayer_Ham}
\end{equation}
The diagonal blocks describe two graphene layers rotated with respect to each other by an angle $\theta$. The Hamiltonians for each of the single graphene layers is given by
\begin{equation}
	h(\theta,\vect k)=\gamma\begin{pmatrix}
	0&f(R(\theta)\vect k)\\
	f^*(R(\theta)\vect k)&0
	\end{pmatrix},
\end{equation}
where $f(R(\theta)\vect k)$ is the geometric factor determined by nearest-neighbor intra-layer hopping, $f(\vect k) = \sum_{\vect \delta} e^{i \vect \delta \cdot \vect k}$, $\gamma=v_F/a_0$ in natural units ($\hbar = c = e = 1$), and $R(\theta)$ is the rotation matrix with rotation axis perpendicular to the sample surface. The momentum shifts $\kappa_{\pm}$ are introduced to fix a common origin for the two layers. When we are interested in the low-energy physics, the geometric factor $f(R(\theta)\vect k)$ can be linearized around the $K,K'$ points.  In Fig. \ref{fig:twisted_fig1}(b) we show the moir\'e Brillouin zone (MBZ).  

The interlayer hopping matrix 
\begin{align}\label{eq:tunn_1}
 	&T(\vect x)=\sum_{i=-1}^1 e^{-i\vect b_i\cdot \vect x} T_i,\\
 	&T_i=w_0\mathbb{1}_2+w_1\left(\cos\left(\frac{2\pi n}{3}\right)\sigma_1+\sin\left(\frac{2\pi n}{3}\right)\sigma_2\right),
 	\label{eq:tunn_2}
\end{align}
where $\vect b_0=(0,0)$, and $\vect b_{\pm 1}= k_\theta\left(\pm \sqrt{3}/2,3/2\right)$ are the reciprocal lattice vectors, captures the dominant tunneling processes between the layers and gives rise to the moir\'e pattern. The parameter $w_1$ describes relaxation effects ~\cite{PhysRevB.96.075311,fleischmann2019moir} and changes in the interlayer-lattice constants~\cite{PhysRevB.99.205134}. 

The symmetries of the continuum model Eq. (\ref{twist_bilayer_Ham}) include a three-fold rotational symmetry centered at the AA region ($C_3$), a two-fold rotation $C_2$ about the same axis composed with time-reversal symmetry $T$ (accounting for the two valleys), and mirror symmetry $M_y$ ~\cite{PhysRevB.99.035111,Balents2019,PhysRevX.8.031089}. In this section, we consider the parameters $\gamma=v_F/a_0=2.36$ eV, and $a_0 = 2.46\mbox{ \normalfont\AA}$. Figure \ref{fig:twisted_fig1}(c) shows a typical band structure for twisted bilayer graphene for  $\theta=1.05^\circ$, close to the magic angle. We highlight the flat bands in red.

\subsubsection{Light in free space}

Previous works have investigated the effects of light on TBG. In Ref. \cite{Topp_2019}, Gabriel E. Topp et al.  consider TBG with twist angles above the magic angle, and 
show that circularly polarized light can induce topological transitions. In Ref. \cite{katz2019floquet}, Or Katz et al.  show that light beams in the visible-infrared range can the emergence of topological flat isolated Floquet-Bloch bands.  In Ref. \cite{li2019floquetengineered}, Yantao Li et al.  show that Floquet flat bands with non trivial topology can be generated with circularly polarized UV laser light. Here, we focus in the low-frequency regime.

We assume that circularly polarized light is applied to the twisted sample at normal incidence. Using the Peirls substution for hoppings $t_{ij}\to e^{i\int_{R_i}^{R_j}\vect A d\vect r} t_{ij}$, the time-dependent Hamiltonian  becomes
\begin{equation}
\begin{aligned}
&H_{\vect k}(\vect x)=\begin{pmatrix}
h(-\theta/2,\vect k(t)-\kappa_-)&T(\vect x)\\
T^\dag(\vect x)&h(\theta/2,\vect k(t)-\kappa_+)
\end{pmatrix}
\end{aligned},
\label{twist_bilayer_Ham_time_free}
\end{equation}
where $H(\vect x, t+2\pi/\Omega) = H(\vect x, t) $, $ k_x \to \tilde k_x(t)  = k_x - A \cos (\Omega t)$, and $k_y \to \tilde k_y = k_y - A \sin (\Omega t)$. The tunneling sector is not modified, since the interlayer tunneling is dominated by processes between atoms that are localized on top of each other and define the moir\'e superlattice. 

We are interested in deriving an effective Floquet model for twisted bilayer graphene using the method described in Sec. \ref{sec:selfconsistentlow}. We consider first the effect low-energy bands at small angles and for weak drives. The time-dependent Hamiltonian within these approximations has the form
\begin{equation}
 H(t)=  H_L+\mathcal Pe^{-i\Omega t}+\mathcal P^\dag e^{i\Omega t},
\label{eq:monocromatic2}
\end{equation}
where the monochromatic operator is $\mathcal{P} = T^{-1}\int_0^T ds H(\vect x , s) e^{i \Omega s}$ and $H_L$ the linearized momenta approximation to Eq. \eqref{twist_bilayer_Ham}. According the theory outlined in Sec. \ref{sec:selfconsistentlow}, the the effective self-consistent time-independent Hamiltonian is~\cite{Vogl2020_effham} 
\begin{equation}
H_{\mathrm{eff}} (\epsilon) \approx H_L+P\frac{1}{\epsilon-H_L-\Omega }P^\dag+P^\dag\frac{1}{\epsilon-H_L+\Omega} P.
\label{eq:truncfrac}
\end{equation}

The Hamiltonian Eq.\eqref{eq:truncfrac} is valid for arbitrary frequency. In particular, in the high-frequency regime we obtain $H_{\mathrm{eff}}=H_L+H_\Omega$, with $H_\Omega=-\Delta \tau_0 \otimes \sigma_3$, $\Delta= (A \gamma a_0)^2/\Omega$, and $\sigma_i$, $\tau_i$ are the Pauli matrices in pseudo-spin and layer space, respectively. This result was first derived in Refs.~\cite{katz2019floquet,li2019floquetengineered}, and shows that the main effect of high-frequency light is to create a topological gap due to time-reversal symmetry breaking $\mathcal{T}$.  

In the low-frequency regime, and for small-enough twist angles defined by the condition $\min\lVert T(\vect x)\rVert\gg \lVert h(\vect k)\rVert$, where $\lVert .\rVert$ is a matrix norm, we obtain $H_{\mathrm{eff}}=H_0+H_\Omega+\mathcal{O}\left(\left(\frac{A}{k_D}\right)^3,\left(\frac{A}{k_D}\right)^2\frac{k_\theta}{k_D}\right)$ where

\begin{align}\nonumber
 H_{\Omega}(\vect x)& =  V(\vect x,\Omega) \mathcal \tau_0 \otimes \sigma_0 +  U(\vect x,\Omega)  \tau_3 \otimes \sigma_0  + \frac{1}{2}\Delta_1(\vect x,\Omega)  (\tau_0+\tau_3) \otimes \sigma_3 + \frac{1}{2}\Delta_2(\vect x,\Omega)  (\tau_0-\tau_3) \otimes \sigma_3\\ & +   \delta w_0(\vect x,\Omega) \tau^+ \otimes \sigma_0 +  \delta w^*_0(\vect x,\Omega) \tau^- \otimes \sigma_0 + \beta(\vect x,\Omega)\tau^+ \otimes \sigma_3+ \beta^*(\vect x,\Omega)\tau^- \otimes \sigma_3.
\label{eq:low-freq-corr}
\end{align}

\begin{figure}[t]
	\begin{center}
		\subfigure{\includegraphics[width=8.50cm]{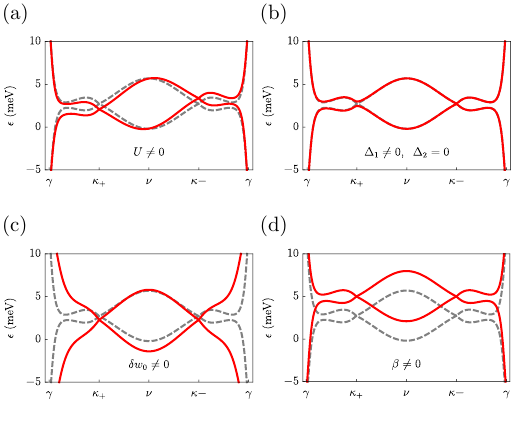}}		
		\caption{(Color online) Sketch of the individual effects of the new term generated by low-frequency and low-intensity circularly polarized light on the  TBG quasienergies. The parameters used are $w_0=w_1 = 110 $~meV, and $\theta=1.2^\circ$. The gray dashed curves correspond to the static case, while the red curve indicates the effect introduced by the non-zero perturbation introduced by light. Figure reprinted from Ref. \cite{vogl2020effective}.
		}
		\label{fig:fig_loweffec_fig2}
	\end{center}
\end{figure}

The matrix structure of the Hamiltonian Eq. \eqref{eq:low-freq-corr} reveals a more reach structure compared with the effective Hamiltonian in the high-frequency regime. $V( \vect x, \Omega ) \mathcal \sigma_0 \otimes \tau_0$ corresponds to an overall position-dependent potential. $U(\vect x, \Omega)$ is a position-dependent interlayer bias which breaks mirror symmetry $M_y$. $\Delta_{1/2}(\vect x, \Omega)$ breaks $M_y$, and $C_2 \mathcal T$ symmetry, which protects the linear band crossing, leading to the opening of a gap at the $\kappa_{\pm}$ points in the mBZ. $\delta w_0(\vect{x}, \Omega)$ introduces a correction to the tunneling amplitude $w_0$ and effectively renormalizes the Fermi velocity at the $\kappa_{\pm}$ points. Finally, $\beta(\vect{x}, \Omega)$ can be interpreted as a pseudo-spin dependent tunneling term and breaks both $C_2 \mathcal T$ and $M_y$. A summary of the individual effects of the new terms are shown in Fig. \ref{fig:fig_loweffec_fig2}. 

In the large-angle limit defined by the condition $\sqrt{(\vect k-\kappa_+)^2+(\vect k-\kappa_-)^2}\gg 3\frac{w_1}{\hbar v_F}$, we find that  $H_{\mathrm{eff}}=H_0+H_\Omega+\mathcal{O}\left(\left(\frac{A}{k_D}\right)^3,\left(\frac{A}{k_D}\right)^2\frac{w_{1,2}}{\gamma}\right)$, where the leading correction is given by

\begin{equation}
	H_{\Omega}(\vect k) =  V(\vect k,\Omega) \mathcal \tau_0 \otimes \sigma_0 +  U(\vect k,\Omega)  \tau_3 \otimes \sigma_0  + \frac{1}{2}\Delta_1(\vect k,\Omega)  (\tau_0+\tau_3) \otimes \sigma_3 + \frac{1}{2}\Delta_2(\vect k,\Omega)  (\tau_0-\tau_3) \otimes \sigma_3
\end{equation}

where $\Delta_{1/2}(\vect k,\Omega)$, $U(\vect k,\Omega)$, and $V(\vect k,\Omega)$ are continuous function of momentum. A summary of each term's effect on the symmetries of the systems is presented in Table \ref{tab:symmetries}.

\begin{table}[]
\centering
\begin{tabular}{|c|c|c|c|}
\hline
                           & \multicolumn{1}{c|}{$C_2 T$}      & \multicolumn{1}{c|}{$C_3$} & \multicolumn{1}{c|}{$M_y$} \\ \hline
$U$                        & $\checkmark$                      & $\checkmark$               & $\text{x}$                 \\ \hline
$U(\vect x)$               & $\text{x}$                        & $\text{x}$                 & $\text{x}$                 \\ \hline
$U(\vect k)$               & \multicolumn{1}{c|}{$\checkmark$} & $\checkmark$               & $\checkmark$               \\ \hline
$\Delta$                   & $\text{x}$                        & $\checkmark$               & $\text{x}$                 \\ \hline
$\Delta(\vect x)$          & $\text{x}$                        & $\text{x}$                 & $\text{x}$                 \\ \hline
$\Delta(\vect k)$          & $\text{x}$                        & $\checkmark$               & $\text{x}$                 \\ \hline
$\delta \omega_0$          & $\checkmark$                      & $\text{x}$                 & $\checkmark$               \\ \hline
$\delta \omega_0(\vect x)$ & $\checkmark$                      & $\text{x}$                 & $\checkmark$               \\ \hline
$\beta$                    & $\checkmark$                      & $\text{x}$                 & $\checkmark$               \\ \hline
$\beta (\vect x)$          & $\text{x}$                        & $\text{x}$                 & $\text{x}$                 \\ \hline
\end{tabular}
\caption{Effect of the light-induced terms in twisted bilayer graphene. The spatial or momentum dependence relates to the small- and large-twist angle regime. A checkmark means that the symmetry is preserved, while a cross that symmetry is broken.}
\label{tab:symmetries}
\end{table}

Low-frequency light can induce a range of symmetry breaking process, which can be controlled by tuning the properties of the incident pulse such as frequency, amplitude, and phase.

\subsubsection{Light confined into a waveguide}

\begin{figure}[]
\begin{center}
	\includegraphics[width=0.6\linewidth]{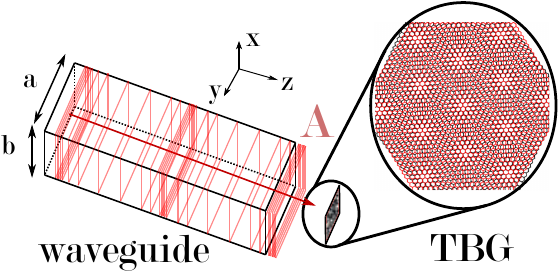}
	\caption{Sketch of twisted bilayer graphene placed at the exit of a rectangular waveguide with cross section $ab$. $\vect A$ is the longitudinal vector potential. Reprinted from Ref. \cite{vogl2020tuning}}
	\label{fig:placeholdertwisted}
\end{center}	
\end{figure}

In this section, we consider the effect of light confined into a waveguide in the electronic states of twisted bilayer graphene. A sketch of the system considered in this section is shown in Fig. \ref{fig:placeholdertwisted}. When the light pulse travels through the waveguide, the boundary conditions in the magnetic and electric field imposed by the metallic walls admit a vector potential of the form
\begin{equation}
 \vect A = \hat z A \sin\left(m\pi x/a\right)\sin\left( n \pi y / b \right) \mathrm{Re}(e^{-ik_z z-i\Omega t}),
 \end{equation}
where  $k_z=\sqrt{k^2-(m\pi/a)^2-(n \pi/ b)^2}$ is the wave number in the $z$-direction, $k^2=\Omega^2 \mu \varepsilon$, $\mu$ is the permeability constant of the insulator inside the waveguide, $m,n\in \mathbb{Z}$ characterize the transverse modes and $\varepsilon$ is the dielectric constant. In the limit where the sample is small compared with the waveguide cross section $ab$, we can assume that the vector potential is position independent: $\vect A=A \mathrm{Re}(e^{-ik_z z-i\Omega t}) \hat z$. 

Under the influence of this drive, only the tunneling sector of the Hamiltonian acquires time dependence. The Hamiltonian is given by
\begin{equation}
\begin{aligned}
&H_{\vect k}(\vect x)=\begin{pmatrix}
h(-\theta/2,\vect k-\kappa_-)&T(\vect x,t)\\
T^\dag(\vect x,t)&h(\theta/2,\vect k-\kappa_+)
\end{pmatrix},
\end{aligned}
\label{twist_bilayer_Ham_time_waveguide}
\end{equation}
where the time-dependent interlayer hopping matrix is introduced via the time-dependence of $w_0 \rightarrow w_0e^{-ia_{AA}}$ and $w_1 \rightarrow w_1e^{-i a_{AB} A\cos(\Omega t)}$, where
%
%
$a_{AA}=3.6\mbox{ \normalfont\AA}$ is the distance of the graphene layers in $AA$-stacked regions and $a_{AB}=3.4\mbox{ \normalfont\AA}$ the corresponding quantity for $AB$ stacking \cite{PhysRevB.99.205134}. 

Now we study the effective Floquet Hamiltonians. In the high-frequency regime ($W< \Omega$, where $W$ is the bandwidth of the system), to leading order in $\Omega^{-1}$ we obtain a renormalization of the hopping amplitudes as 
\begin{align}
	w_1 \to & \tilde w_1=J_0\left(\left| a_{AB} A\right|\right) w_1,\\
	w_0\to & \tilde w_0= J_0\left(\left| a_{AA} A\right|\right)w_0,
\end{align}
where $J_0$ is the zeroth Bessel function of the first kind. This result does not depend on the details of the layer sector of the Hamiltonian, and can be applied to other low-dimensional heterostructures to weaken their interlayer tunneling\cite{Cheng_2019,Liu_2014,wu2019anomalous,PhysRevB.96.035442,shang2019artificial,Abdullah_2017}. Notice that in this regime longitudinal light does not break time-reversal symmetry, allowing control of the Fermi velocity while preserving the linear band crossing at the $\kappa_+$ point in the MBZ.


In the chiral limit ($w_{0}=0$), Tarnopolsky \textit{et al. } Ref.~\cite{PhysRevLett.122.106405} found perfectly flat bands appearing at $\alpha_1 \approx0.586$ or $\theta_1\approx 1.09^\circ$ where $\alpha=w_1/(2v_Fk_D\sin(\theta/2))$. Further flat bands appear at smaller angles $\alpha_n=\alpha_1+n\Delta\alpha$ with $\Delta \alpha = 3/2$ and $n\in \mathbb{N}$. In the driven case, the magic angles appear at
\begin{equation}
	\theta_n=\frac{w_1J_0\left(\left| a_{AB} A\right|\right)}{v_F k_D \alpha_n}.
	\label{eq:magic_angle_driven}
\end{equation}
The accuracy of Eq. \eqref{eq:magic_angle_driven} was verified numerically employing a diagonalization of the extended space Hamiltonian. The potential flexibility in the twist angle depends on argument of the Bessel function, $\eta \equiv e a_{A B} E /(\hbar \Omega)$. In a pump-probe setup with pump drive frequency $f=\Omega /(2 \pi)=650 \mathrm{THz}$, and peak  electric field strength $E=15 \mathrm{MV} / \mathrm{cm},\eta \approx 0.19,$ which leads to $\theta_{F} / \theta \approx 0.99$. For $E=$ $25 \mathrm{MV} / \mathrm{cm}$, $\theta_{F} / \theta=0.975$. These peak electric fields have been used on graphene before . For example, peak electric fields of up to $\sim 30 \mathrm{MV} / \mathrm{cm}$ with frequencies in the near-IR regime $\sim 375$ $\mathrm{THz}$ led to light-field-driven currents ~\cite{PhysRevLett.121.207401,Higuchi2017}. 

\begin{figure}[]
\begin{center}
	\includegraphics[width=1\linewidth]{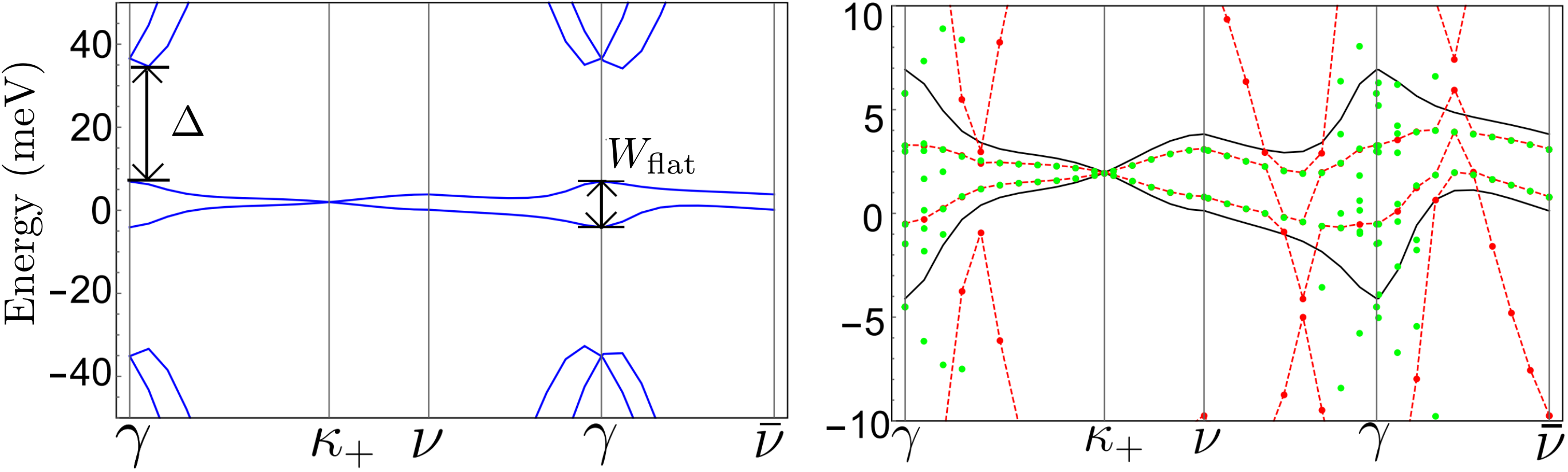}
	\caption{(Color online) (a) Band structure of twisted bilayer graphene for $w_{0}=88 \mathrm{meV},$ and $w_{1}=110 \mathrm{meV}.$ The gap between the central flat bands and the continuum is indicated by $\Delta$. (b) Central Floquet quasienergies for $\Omega=20 \mathrm{meV}$ and $E \sim 0.7 \mathrm{MV} / \mathrm{cm}$. The black lines are the static case, and the coinciding red and green points indicate convergence. Reprinted from Ref. \cite{vogl2020tuning}}
	\label{fig:low-freq-waveguide}
\end{center}	
\end{figure}

Now we consider the low-frequency regime, defined by $\Omega < W$. In particular we consider only off-resonant drives with $\Delta > \Omega > W_{\rm{flat}}$, where $\Delta$ is the energy gap between the low-energy flat bands and the continuum and $W_{\rm{flat}}$ is the bandwidth of the flat bands, as shown in Fig. \ref{fig:low-freq-waveguide}(a). In this regime, we restore to fully numerical methods.

In Figure \ref{fig:low-freq-waveguide}(b), we plot the converged low-quasienergy bands along a high-symmetry path in the MBZ for $w_{0}=88 \mathrm{meV},$  $w_{1}=110 \mathrm{meV}$, $E=0.7 \mathrm{MV} / \mathrm{cm}$, and $\Omega = 20 $~meV. We find that the Fermi energy has been reduced compared to the static case, which indicates that the flat bands are shifted to larger angles. The magic angle is found at $\theta \approx 1.12$ Degrees, compared with $\theta \approx 1.10$ Degrees for the static case. 

Remarkably, this shows that depending on driving frequencies we can either increase or lower the magic angle. Since the appearance of the strongly correlated phases depends on the twist angle, applying light could be employed to tune in and out of these phases. 

\subsection{Twisted double bilayer graphene}

In this section we consider the effect of light on twisted double bilayer graphene (TDBG). We consider both AB/AB and AB/BA stacks. In contrast with twisted bilayer graphene, TDBG allows for independent control of the quasienergy gaps near the $K$ and $K'$ points of the BZ and the $\kappa_\pm$ points of the MBZ. 

\subsubsection{Static system}
\label{sec:tdbg_static}
\begin{figure}[t]
	\begin{center}
		\subfigure{\includegraphics[width=14.00cm]{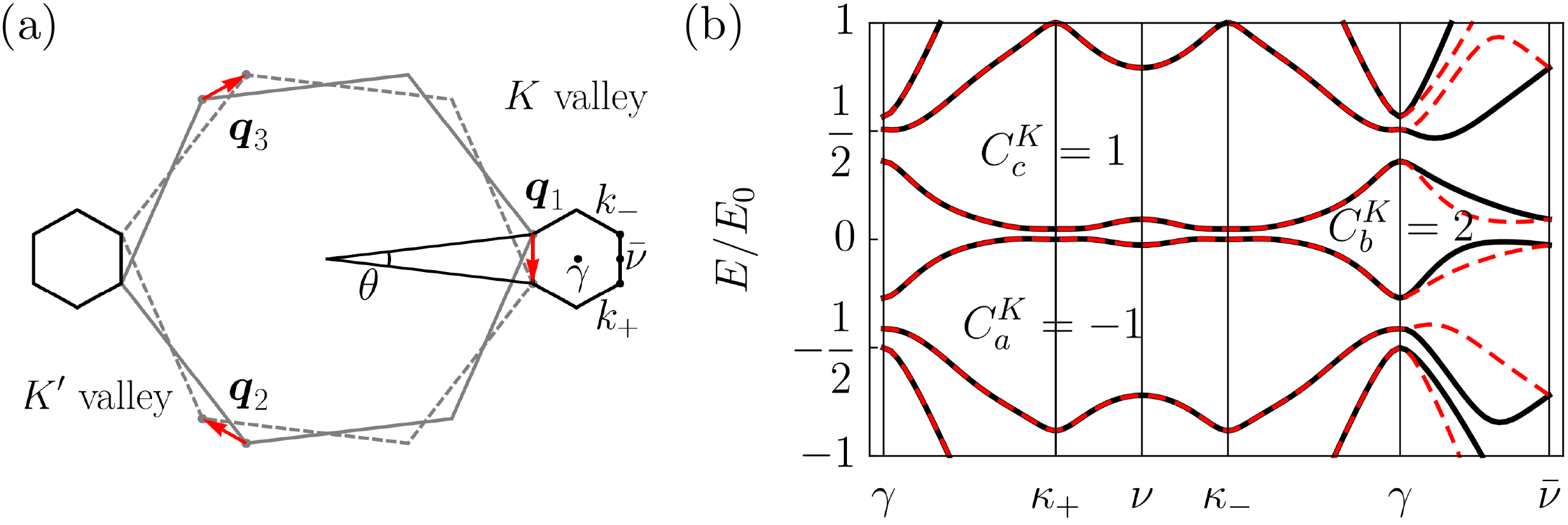}}		
		\caption{(Color online) (a) AB/BA TDBG band structure for $\theta = 1.4^\circ$, $\Delta=0$ and $\gamma_{3/4} = 0$ along a high symmetry path in the mBZ. The black solid (red dashed) lines correspond to the spectrum near the $K$ ($K'$) point. The Chern numbers in the gaps labeled $a$, $b$, and $c$ are indicated for the $K$ point. Time reversal symmetry imposes $C^{K'}=-C^{K}$. The energy scale is $E_0=100$~meV. (b) Moir\'e Brilloiun zone (MBZ). Figure reprinted from Ref. \cite{rodriguezvega_2020a}}
		\label{fig:fig2}
	\end{center}
\end{figure}

The continuum-limit Hamiltonian for static TDBG near the $K$ point with AB/AB ($s=s'=1$) [AB/BA ($s=-s'=1$)] stacking patterns is\cite{PhysRevB.99.235417,PhysRevB.99.235406,Lee2019}
\begin{align}\nonumber
H_{ss'}(\vect k,\vect x)  & = \tau_u \otimes h_s(-\theta/2, \vect k-\kappa_-) \\ \nonumber &  +  \tau_d \otimes h_{s'}(\theta/2,\vect k-\kappa_+) \\
&+ \tau^+ \otimes \lambda^- \otimes T(\vect x)  + \tau^- \otimes \lambda^+ \otimes T^{\dagger}(\vect x) ,
\label{twist_bi-bilayer_Ham}
\end{align}
where $\tau_u  = \left(\mathbb 1 + \tau_3 \right)/2$, $\tau_d  = \left(\mathbb 1 - \tau_3 \right)/2$, $\tau_{\pm}  = \left(\tau_1 \pm i \tau_2 \right)/2$, and $\tau_i$ and $\lambda_i$ are Pauli matrices in top/bottom bilayer and layer space, respectively. Here, $\sigma_k$ are Pauli matrices or identity operators in pseudospin space. The rotated bilayer graphene Hamiltonian~\cite{jung2014} $h_s(\theta, \vect k)$ includes trigonal warping and  effects and particle-hole symmetry breaking terms. The interlayer hopping matrix $T(\vect x)$ is the same as for TBG, since we neglect direct tunneling contributions between layers that are not adjacent to one another.

The Hamiltonian near the $K'$ valley can be obtained by applying a time reversal operation $\mathcal T$ to the Hamiltonian at the $K$ valley~\cite{Balents2019}. Before studying the time dependent case it is worthwhile to summarize various symmetry properties of static TDBG.  In addition to time-reversal symmetry $\mathcal T$,  AB/AB TDBG possesses $C_{3z}$ rotational symmetry, and mirror symmetry $M_x: y, k_y \to -y,-k_y$ in the absence of an applied static electric field. The AB/BA TDBG possesses $C_{3z}$, mirror symmetry $M_y: x,k_x \to -x,-k_x$ (which switches the valleys), and $M_y \mathcal T$~\cite{PhysRevB.99.075127,Lee2019,PhysRevB.99.235406}. 


TDBG can be topologically non-trivial, dependending on the stacking pattern. Time reversal symmetry imposes the condition $C^K_{n} = -C^{K'}_{n}$, where $C^{K(K')}_{n}$ is the Chern number for band $n$ in the vicinity of the $K(K')$ point. For AB/AB stacking, $M_y$ symmetry  implies $C^{K/K'}_{n}=0$ for each $n$~\cite{Lee2019}. On the other hand, AB/BA TDBG possesses non-zero $C^{K(K')}_{n}$ leading to a Hall valley insulating phase.

In the next section, we will explore the effect of longitudinal and circularly polarized light on TDBG. 

\subsubsection{Driven system in free space}
\label{sec:driv_free}

The time-dependent Hamiltonian  $ H_{ss^\prime}(t)  \equiv H_{ss^\prime}(\vect k(t),\vect x) $ arises from the minimal coupling \cite{PhysRevB.101.205140}
and $k_x(t)  = k_x - A \cos (\Omega t)$, and $k_y(t) = k_y - A \sin (\Omega t)$.  In the high-frequency regime, a van Vleck expansion~\cite{Eckardt_2015} leads to the effective Hamiltonian $H^s_{\text{Vv}} = H_s^{(0)} + \delta H_{s,\text{vV}}$, where $H_s^{(0)}$ is the zeroth order Hamiltonian, 
\begin{align}
\delta H_{ss',\text{vV}} = -(\Delta_{\text{vV}}-\Delta^{(3)}_{\text{vV}}) \mathbb{1} \otimes \mathbb{1} \otimes  \sigma_3 - (\Delta^{(4)}_{\text{vV}} - \Delta^{(3)}_{\text{vV}}) 
& \left( s \tau^u  \otimes  \lambda_3  \otimes \mathbb{1} + s' \tau^d \otimes  \lambda_3  \otimes \mathbb{1} \right),  
\label{eq:HvV}
\end{align}
$\Delta_{\text{vV}}= \xi (v_F A)^2/\Omega$,  $\Delta^{(4)}_{\text{vV}}=\xi (v_4 A)^2/\Omega$, and $\Delta^{(3)}_{\text{vV}}=\xi (v_3 A)^2/(2\Omega)$, where $\xi=1$($\xi=-1$) near the $K$($K'$) valley. 

The new terms induced by the light can open up a gap in the quasienergy spectrum and can lead to topological transitions, depending on the stacking configuration. For example, AB/AB TDBG is a trivial insulator for in the presence of mirror symmetry. $\Delta_{\text{vV}}$ induced by circularly-polarized light leads to a transition into a Chern insulator with Floquet topological bands due to time-reversal symmetry breaking. On the other hand, in equilibrium, AB/BA TDBG is a valley Chern insulator. Circularly polarized light also leads to a transition into a Floquet Chern insulating phase. This selective gap engineering could be employed to generate valley-polarized currents in AB/BA TDBG. 

To access the intermediate-frequency regime, we employ a modified rotating frame transformation~\cite{vogl2020effective} and a time average. The resulting effective Floquet Hamiltonian is 
\begin{align}
H^{ss'}_{F} = R^{\dagger}(\theta) \left( \bar H_{ss'} + \delta H_{F} \right) R(\theta),
\label{eq:rotated_ham_mid_freq}
\end{align}
where $R(\theta)$ is a twist-angle dependent unitary transformation, and $ \delta H_{F} =  \Delta_F  \mathbb{1} \otimes \mathbb{1} \otimes \sigma_3 $, with $\Delta_F = A J_1(2 \sqrt{2} A/\Omega)/\sqrt{2}$, where $J_n(z)$ correspond to the $n$-th Bessel function of the first kind. $\bar H_{ss^\prime}$ shares the same structure with the static Hamiltonian with the following renormalized parameters: $\tilde \gamma_0 = J_0(2 A/\Omega) \gamma_0 = J_0(2 A/\Omega) v_F/a_0 $, $\tilde \delta^{\pm}_s = \delta J_0(2\sqrt{2} A/\Omega) ( 1\pm s)/2$, and $\tilde t_s= \gamma_1 J_0(2 A/\Omega) (\sigma_1-i s\sigma_2) /2$. None of these effects are captured in a leading-order van Vleck expansion, and its challenging to capture the functional form simply by computing higher-order terms. 

The interlayer coupling renormalizes to
\begin{equation}
\begin{aligned}
&\tilde T(\vect x)=\sum_{n=-1}^1 e^{-i \vect Q_n \cdot \vect x}(\tilde T_n-i\omega_\theta\sigma_3),
\end{aligned}
\end{equation}
where the tunneling amplitudes in the static $T_n$ matrix are renormalized to $\tilde \omega_1=J_0(2 A/\Omega)\omega_1$,  $\tilde \omega_0=\omega_0+\sin ^2(\theta/2 ) \left(J_0\left(\frac{2 \sqrt{2} A}{\Omega }\right)-1\right)\omega_0$. The the term 
\begin{equation}
\omega_\theta=\frac{1}{2} \sin ( \theta ) \left(J_0\left(\frac{2 \sqrt{2} A}{\Omega }\right)-1\right)\omega_0,
\end{equation}
is an angle-dependent coupling not present in equilibrium. 

The effective Hamiltonian Eq. \eqref{eq:rotated_ham_mid_freq} is accurate up to frequency and driving strength regimes where the van Vleck approximation breaks down. For example, for a driving frequency $\Omega/W=2$ one can describe gaps with errors below $10\%$ up to driving strengths $a_0A \approx 1$, in contrast to the van Vleck approximation only manages to do so until $a_0A \approx 0.45$. Therefore, the implementation of an improved transformation into a rotating frame can enhance the range of validity of effective Floquet Hamiltonians when it comes to driving strengths. Accurate effective Hamiltonians are of significance for the study of driven system combined with computationally challenging additional effects such as disorder.

\begin{figure}[h]
	\begin{center}
		\subfigure{\includegraphics[width=9.50cm]{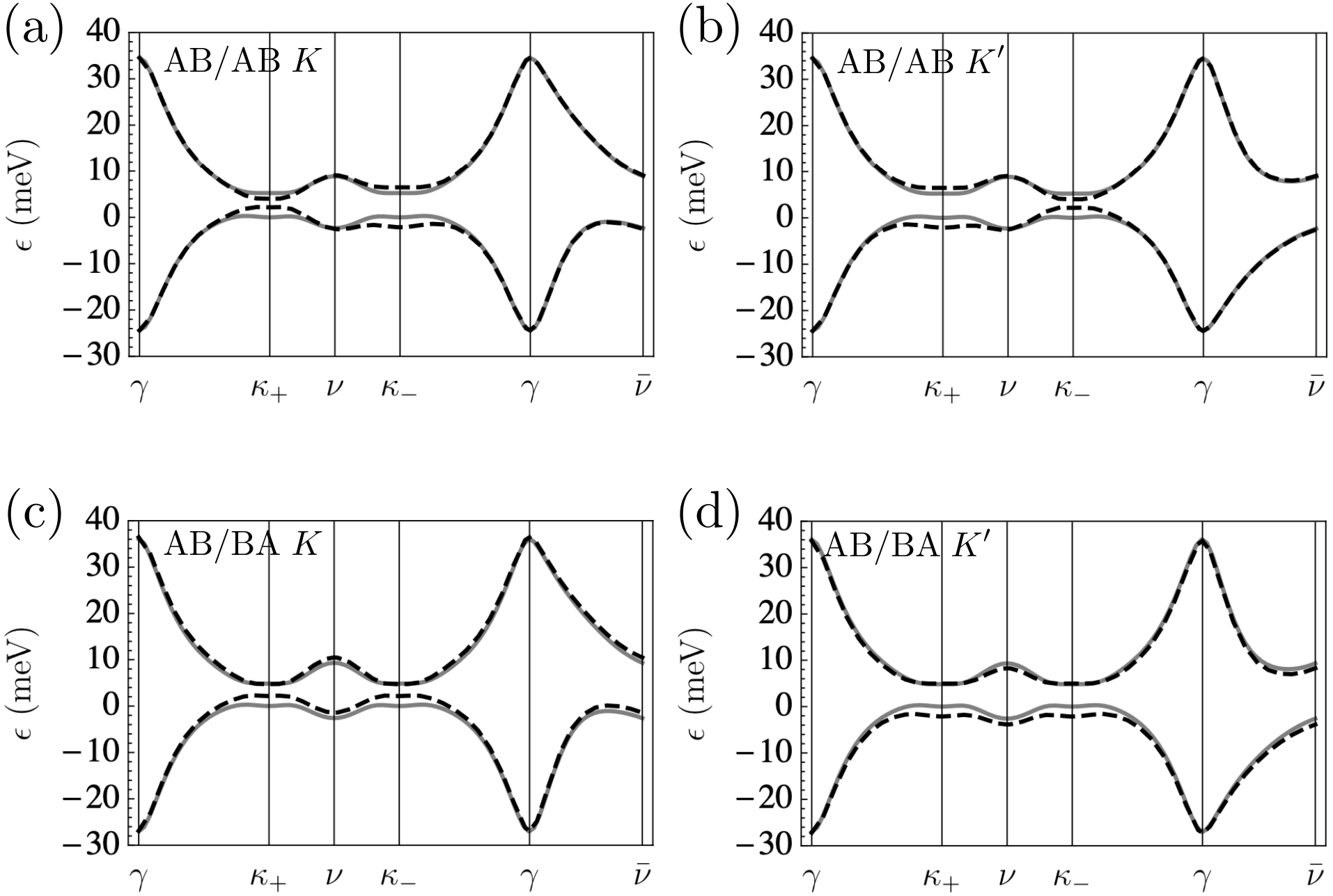}}		
		\caption{(Color online) Quasienergies near the $K$ and $K'$ points for (a-b) AB/BA and (c-d) $AB/BA$ stacked TDBG with $\theta=1.4^{\circ}, \Omega / W=2,$ and $a_{0} A=0.04$.  The gray curves indicate the equilibrium energies.  Figure adapted from Ref. \cite{rodriguezvega_2020a}}
		\label{fig:fig_tdbg_bands}
	\end{center}
\end{figure}

\subsubsection{Driven system in a waveguide} 
\label{sec:waveguide_double-bilayer}

For light confined into a waveguide, the time dependence enters through the tunneling sectors, leading to the time-dependent Hamiltonian
\begin{align}\nonumber
H_{ss'}(\vect k,\vect x,t) &  = \tau_u \otimes h_s(-\theta/2, \vect k-\kappa_-,t)     +  \tau_d \otimes h_{s'}(\theta/2,\vect k-\kappa_+,t) \\
&+ \tau^+ \otimes \lambda^- \otimes T(\vect x,t)  + \tau^- \otimes \lambda^+ \otimes T^{\dagger}(\vect x,t) ,
\label{twist_bilayer_Ham_waveguide}
\end{align}
where 
\begin{align}
  h_s(\theta,\vect k,t)&= \left(\begin{array}{cc|cc}
       \Delta_1 + \delta^-_s & \gamma_0 f(R_\theta \vect k)    &     \multicolumn{2}{c}{\smash{\raisebox{-.5\normalbaselineskip}{$t_s(\vect k,t)$}}} \\
       \gamma_0 f^*(R_\theta\vect k)&\Delta_1+ \delta^+_s &           &   \\
      \hline \\[-\normalbaselineskip]
     \multicolumn{2}{c|}{\smash{\raisebox{-.5\normalbaselineskip}{$t^{\dagger}_s(\vect k,t)$}}} &     \Delta_2 + \delta^+_s& \gamma_0 f(R_\theta\vect k)\\
       &        &  \gamma_0 f^*(R_\theta\vect k) & \Delta_2+ \delta^-_s
    \end{array}\right),
    \label{eq:bilayer_waveguide}
\end{align}
\begin{align}
	&t_{+}(\vect k,t)=
	 \begin{pmatrix}
	 - \gamma_4 f(R_\theta \vect k) & - \gamma_3 f^*(R_\theta\vect k) \\
	 \gamma_1 & - \gamma_4 f(R_\theta \vect k)
	\end{pmatrix} e^{- i a_{AB}A(t)} .
\end{align}
The interlayer hopping matrix acquires a time dependence according to
 \begin{align}
 	&T(\vect x,t)=\sum_{i=-1}^1 e^{-i\vect Q_i \cdot \vect x} T_i(t),\\ \nonumber
 	&T_i= e^{- i a_{AA}A(t)}w_0\mathbb{1}_2+ e^{- i a_{AB}A(t)} w_1   \left(\cos\left(\frac{2\pi i}{3}\right)\sigma_1+\sin\left(\frac{2\pi i}{3}\right)\sigma_2\right).
 	\label{eq:tunn_2_waveguide}
\end{align}

In the high-frequency regime, we employ again a first-order van Vleck expansion. To leading order, the effective Hamiltonian is given by the averaged Hamiltonian $H_0$ with renormalized parameters 
\begin{equation}
	\begin{aligned}
	&(w_1,\gamma_{1,3,4})\to (w_1,\gamma_{1,3,4})J_0(a_{AB}A)\\
	&w_0\to w_0J_0(a_{AA}A)
	\end{aligned},
\end{equation}
where $J_0$ is the zeroth Bessel function of the first kind. The corrections of order $1/\Omega$ vanish  if derivatives $\partial_i T(\vect x,t)$  are neglected. This is justified because all derivatives  in $H_m$ and $m\neq 0$ appear with a pre-factor $\gamma_{3,4}$ that is small and the terms $[H_{-m},H_m]/(2m\Omega)$ are already suppressed by $1/\Omega$.
In the small-angle regime, where $T(\vect x,t)$ varies slowly in real space this approximation becomes even better because then the corrections that would arise have an additional small factor $\theta$. The renormalized parameters can lead to a renormalization of the quasienergies, without breaking symmetries that could be required for some applications. In the next section, we consider a system with spin-orbit coupling which brings into play topological band-inversion effects, and discuss the effects of light irradiation.

\subsection{Twisted transition metal dichalcogenides}

\begin{figure}[]
	\begin{center}
		\includegraphics[width=14.0cm]{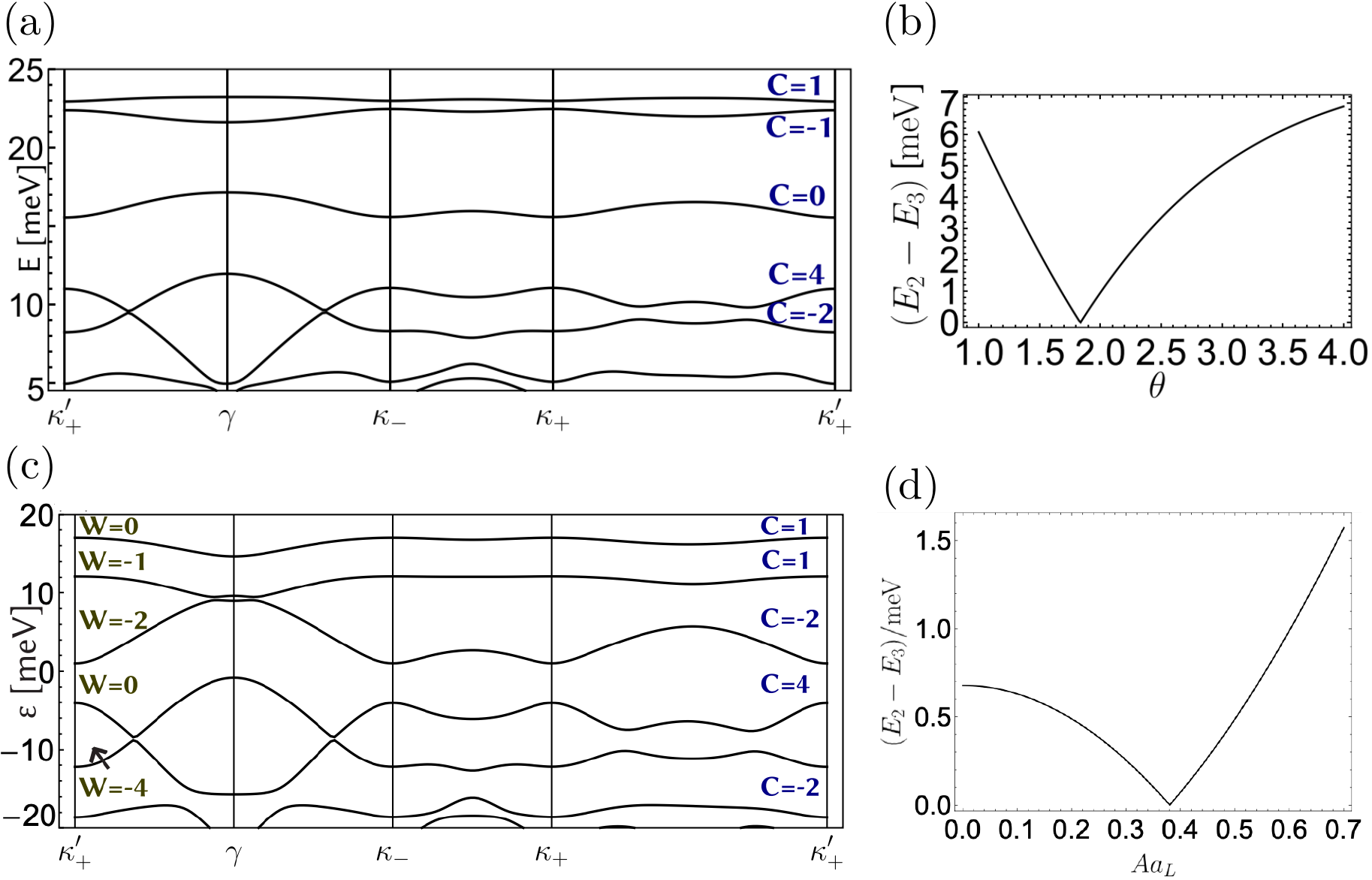}
		\caption{(Color online) (a) Energy bands along a high-symmetry path in the mBZ, along with the band Chern number for $\theta = 1.2^\circ$. (b) Gap between the second and third band (from top to bottom) as a function of twist angle. (c) Quasienergy bands for $\theta = 1.96^\circ$ and the driving frequency $\Omega=0.7$eV. (d) Quasienergy gap between the second and third bands as a function of the laser intensity $A a_L$. 
		Figure adapted from Ref. \cite{vogl2021}. }
		\label{fig:tmd}
	\end{center}
\end{figure}

The low-energy effective Hamiltonian for a twisted TMD homobilayer is~\cite{PhysRevLett.122.086402}
\begin{equation}
\hspace*{-0.245cm}
\mathcal{H}_{\uparrow}( \vect r)[f]=\begin{pmatrix}
f(\vect k-\kappa_+) +\Delta_{1}(\vect r) & \Delta_{T}(\vect r) \\
\Delta_{T}^{\dagger}(\vect r) & f(\vect k-\kappa_-) +\Delta_{-1}(\vect r)
\end{pmatrix},
\label{Hd0}
\end{equation}
where $f(\vect k)$ is the bounded approximate low-energy valence band dispersion of a single layer TMD near the $\vect K$ point. $\mathcal{H}_{\uparrow}( \vect r)[f]$ is defined in the basis $\Psi=(\Psi_b,\Psi_t)$, where $\Psi_b$ ($\Psi_t$) corresponds to the bottom (top) layer creation operator with spin up. The off-diagonal interlayer tunneling sector is
\begin{equation}
\Delta_{T}(\vect r) = w (1+ e^{-i \theta\vect G_2 \cdot (\hat z\times \vect r)}+ e^{-i \theta\vect G_3 \cdot (\hat z\times \vect r)}),
\label{Tunneling}
\end{equation}
where $\vect G_n=4\pi/(\sqrt{3}a_0) R_z((n-1)\pi/3)\hat y$ and $R_z$ is a rotation matrix around the $z$ axis. $w$ determines the strength of the interlayer coupling, and we consider small rotations angles $\theta < 10 ^\circ$. Additionally, this model includes an effective position-dependent layer bias
\begin{equation}
	\Delta_l=2V\sum_{j=1}^3\cos(\theta\vect G_{2j+1}(\hat z\times \vect r)+l\psi),
	\label{delta_l}
\end{equation}
where $V$ sets the strength of the position-dependent in-plane bias and the index $l=\pm 1$. The model parameters are fixed
$
	(V,w,\psi,a_0)=(8 \text{meV},-8.5 \text{meV},-89.6^\circ,3.47 \text{\AA}),
$
following Ref.\cite{PhysRevLett.122.086402} for the case of MoTe$_2$, where $\psi$ a phase term and $a_0$ is the intra-layer distance between sites. The precise shape of the low-energy valence band dispersion $f(\mathbf k)$ and its corresponding coefficients for this material can be found in \cite{vogl2021}.  In Fig. \ref{fig:tmd}(a) shows the band structure along a high-symmetry path in the mBZ, and in Fig. \ref{fig:tmd}(b) the energy differences between the second and third energy bands. At $\theta \approx 1.8$ degrees, the bands close and re-open, leading to a angle-dependent topological transition as indicated by the band Chern numbers $(C_2 = -1 ,C_3=0) \rightarrow (C_2 = 1 ,C_3 = -2)$. 

Now, we demonstrate that light can induce an analogous gap closing and lead to a non-equilibrium topological transition~\cite{vogl2021}. We consider a twisted TMD sample irradiated with longitudinal light from a waveguide. The time-dependent Hamiltonian is obtained via the replacement $w\to e^{-iAa_L\cos(\Omega t)}w$, as discussed in the previous examples. This is equivalent to an electric potential between layers,  up to a gauge transformation. In the high-frequency regime, a van Vleck expansion to first order leads to the effective Floquet Hamiltonian 
\begin{equation}
\hspace*{-0.245cm}
\mathcal{H}_{\uparrow}=\begin{pmatrix}
\tilde f(R_{-\theta/2}(\vect k-\kappa_+)) +\Delta_{1}(\boldsymbol r) & J_0(a_LA)\Delta_{T}(\boldsymbol r) \\
J_0(a_LA)\Delta_{T}^{\dagger}(\boldsymbol r) & \tilde f(R_{\theta/2}(\vect k-\kappa_-)) +\Delta_{-1}(\boldsymbol r)
\end{pmatrix},
	\label{Hdwaveguide}
\end{equation}
By inspecting the effective Hamiltonian, we conclude that light from a waveguide decreases the strength of the interlayer coupling, leading to an effective change in the twist angle. In Fig.~\ref{fig:tmd}(c), shows the quasienergy bands along with the band Chern numbers and winding numbers. As the intensity of the laser is  increased, the gap between the second and third bands decreases until it closes, leading to a topological transition $(C_2 = 1 ,C_3=-2) \rightarrow (C_2 = -1 ,C_3 = 0)$ (see Fig.\ref{fig:tmd}(d)). In turn, the winding number changes from $W=-2$ to $W=0$ across the transition. Therefore, topological transitions can be engineered in TMDs with light.

We have studied graphene- and transition metal dichalcogenide-based van der Waals heterostructures with Moir\'e superlattices driven by circularly polarized light in free space and confined into a waveguide. We considered the high, intermediate, and low-frequency regime. We showed the high degree of tunability of the quasienergy band structure induced by the combined effects of an interlayer rotation and light. In the next section, we depart from single-particle examples, and we consider a magnetic systems driven with infrared light and discuss its effects in the magnetic order.  

\subsection{Phonon-driven Van der Waals magnets}

In this section, we review the effect of light in a strongly correlated state. In particular, we will consider the effect of low-frequency light in resonance with the phonons in the magnetic order of low-dimensional van der Waals antiferromagnets. We will show that dynamically distorting the lattice can lead to AFM-to-FM transitions. The direct coupling of the laser with the electronic degrees of freedom are not considered here. This effect has been addressed in several works~\cite{Mentink2015,liu2018,Hejazi2019doublonholon,Claassen2017,bukov2016,PhysRevB.100.121110}.

\begin{figure}[]
	\begin{center}
		\subfigure{\includegraphics[width=15.50cm]{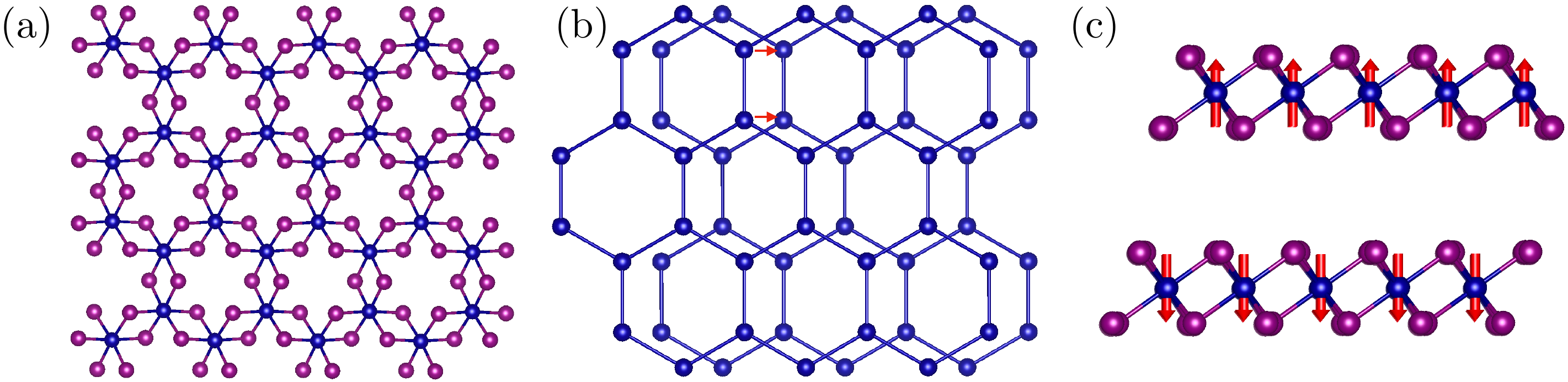}}		
		\caption{(Color online) (a) Lattice structure of one monolayer of CrI$_3$. The Cr atoms (blue) form a hexagonal lattice with surrounding I atoms. The crystal structure for b-CrI$_3$ is obtained by stacking two monolayers and shifting them with respect to each other as shown in (b). Here only Cr atoms are shown for clarity. (c)  b-CrI$_3$ antiferromagnetic order. The lattice structures were created with VESTA~\cite{Momma:db5098}.}
		\label{fig:cri3}
	\end{center}
\end{figure}

As a prototypical example, we consider bilayer CrI$_3$ (b-CrI$_3$). This material presents an antiferromagnetic (AFM) groundstate~\cite{huang2017,seyler2018,klein1218,song2018,sun2019}, with monoclinic crystal structure, as shown in  Fig. \ref{fig:cri3}. Experiments ~\cite{Thiel973,ubrig2020} and first-principles calculations ~\cite{sivadas2018,jang2019,soriano2019,soriano2020} suggest that there is a connection between the magnetic order and the stacking configuration in b-CrI$_3$. The FM phase presents space group R$\bar 3$ (point group S$_6$), while the AFM state presents space group C2/m (point group C$_{2h}$)~\cite{sivadas2018}. 

First, we characterize the phonons in b-CrI$_3$ employing group theory. The primitive unit cell contains $N=16$ atoms, for a total of  $3 N  = 48$ phonon modes. The lattice vibration representation, which characterizes the phonon modes according to their irreducible representation, is  
$
\Gamma_{latt. vib.} = \Gamma^{equiv} \otimes \Gamma_{vec}  =  13 A_g \oplus 11 B_g \oplus 11 A_u \oplus 13 B_u
$.
This indicates that we have 24 Raman active modes (13 with totally symmetric $A_g$ representation and 11 with $B_g$ representation) and 24 infrared active modes~\cite{kroumova2003}. We get the real-space displacements constructing the projection operators~\cite{GroupTheoryDress2008, gtpack1, gtpack2}  
$
\hat P^{(\Gamma_n)}_{kl} = \frac{l_n}{h} \sum_{C_\alpha} \left( D_{kl}^{(\Gamma_n)}(C_\alpha) \right)^* \hat P(C_\alpha)$,
where $\Gamma_n$ are the irreducible representations, $C_\alpha$ are the elements of the group, $D_{kl}^{(\Gamma_n)}(C_\alpha) $ is the irreducible matrix representation of element $C_\alpha$, $h$ is the order of the group, and $l_n$ is the dimension of the irreducible representation. Finally, $\hat P(C_\alpha)$ are $3N\times 3N$ matrices that form the displacement representation. The projection operators $\hat P^{A_g}$ and $\hat P^{B_g}$ indicate that relative shifts $[1 \; 1 \; 0]$ and $[\bar1 \; \bar1 \; 0]$ for the top and bottom layers belong to the totally-symmetric $A_g$ representation. Similarly, $[0 \; 0 \; 1]$ and $[0 \; 0 \; \bar 1]$ for the top and bottom layers are $A_g$ type. Finally,  $[\bar 1 \; 1 \; 0]$ and $[1 \; \bar 1 0]$ are  $B_g$-type. 

\begin{figure}[]
	\begin{center}
		\includegraphics[width=16.0cm]{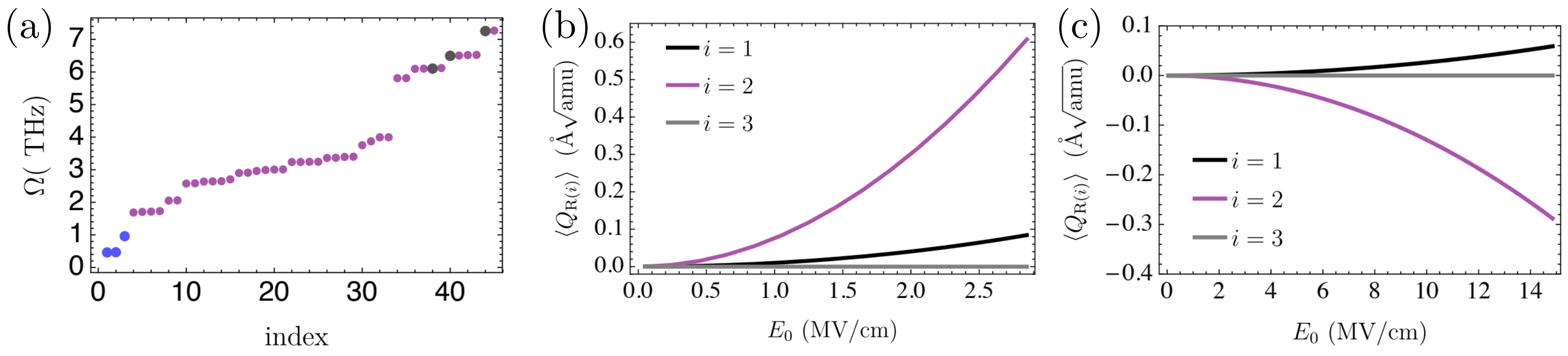}
		\caption{(Color online) (a) b-CrI$_3$ phonon frequencies in the space group $C2/m$.  
		(b) $Q^{(i)}_{\text{R}}$ average displacement due to driving of the infrared $Q_{\text{IR}(A)}$ mode with $\tau=0.3$~ps and $E=4$~MV/cm. In (c), shows the average displacement due to  $\Omega_{\text{IR}}=6.493$~THz with $\tau=0.8$~ps. Figure adapted from Ref. \cite{rodriguezvega_2020a}. }
		\label{fig:qmaxir3}
	\end{center}
\end{figure}

We employ first-principles methods~\cite{Giannozzi_2009,Giannozzi_2017,kresse1996,kresse1996b} to determine the frequencies of these modes at $\Gamma$ point. We show the results in Fig. \ref{fig:qmaxir3}(a). The shear and breathing modes between the layers (blue dots) have frequencies  $\Omega=0.460$~THz ($A_g$ shear mode), $\Omega=0.467$~THz, ($B_g$ shear mode), and  $\Omega=0.959$~THz ($A_g$ breathing mode). 

When the laser irradiates the b-CrI$_3$ sample in resonance with an infrared phonon, and if its amplitude is large enough, phonon-phonon interactions become relevant~\cite{FORST201324,subedi2014,juraschek2017,Juraschek2017b,subedi2017,Juraschek2018,Juraschek2019,Juraschek2020,juraschek2019phonomagnetic,kalsha2018,gu2018,subedi2014,fechner2016,mankowsky2015,mitrano2016}. The symmetry-allowed interaction up to cubic order are 

\begin{align}
V[Q_{\text{IR}},Q_{\text{R}(i)},t] & =  \frac{1}{2}\Omega^2_{\text{IR}} Q_{\text{IR}}^2+ \sum_{i=1}^3 \frac{1}{2}\Omega^2_{\text{R}(i)} Q^{2}_{\text{R}(i)} \nonumber 
+\sum_{i=1}^2 \frac{ \beta_i}{3} Q_{\text{R(i)}}^3 + Q_{\text{IR}}^2 \sum^2_{i=1} \gamma_i Q_{\text{R(i)}} + \delta Q_{\text{R(1)}}^2 Q_{\text{R(2)}}\\ & + \epsilon Q_{\text{R(1)}} Q_{\text{R(2)}}^2+   Q_{\text{R(3)}}^2  \sum^2_{i=1} \zeta_i Q_{\text{R(i)}} + V_D[t,Q_{\text{IR}}]  ,
\label{eq:non-linear-pot_main}
\end{align}

where $Q_{\text{IR}}$ corresponds to the amplitude of an infrared phonon,  and $ Q_{\text{R(i)}}$, $i=1,2,3$ to the amplitudes of the three lowest-frequency Raman modes. The effect of the laser is capture with the time-dependent term~\cite{baroni2001,forst2011}
$
V_D[t,Q_{\text{IR}}]  = \boldsymbol{Z^*} \cdot \boldsymbol{E_0} \sin(\Omega t) F(t) Q_{\text{IR}}
$, 
where $\boldsymbol{E_0}$ is the electric field amplitude, and $\boldsymbol{Z^*}$ is the mode effective charge vector~\cite{gonze1997, baroni2001}. $F(t)=\exp\{ -t^2/(2 \tau^2)\}$ is the Gaussian laser profile, with variance $\tau^2$. In this work, we neglect phonon-damping effects. The coupling constants are determined via first-principles calculations.

The equations of motion for the driven phonons are determined by $\partial^2_t Q_{\text{R}(i)}   = -\partial_{Q_{\text{R}(i)}} V[Q_{\text{IR}},Q_{\text{R}(i)}]$, for $i=1,\cdots,m$, and
$\partial^2_t Q_{\text{IR}}     = -\partial_{Q_{\text{IR}}} V[Q_{\text{IR}},Q_{\text{R}(i)}]$. Here, this set of equations is solved numerically. We consider the IR modes with frequencies $\Omega_{\text{IR}}=6.104$~THz ($Q_{\text{IR}(A)}$). In Fig. \ref{fig:qmaxir3}(b), we plot the averaged displacements $\langle Q_{\text{R}(i)}\rangle$ as a function of the peak electric field $E_0$. The non-zero average of the Raman modes amounts to an effective non-equilibrium lattice distortion. The direction of the distortion $\langle Q_{\text{R}(2)}\rangle$ can be reversed by changing the frequency of the laser, in resonance with $\Omega_{\text{IR}(B)}$ ($\Omega_{\text{IR}}=6.493$~THz) as shown in \ref{fig:qmaxir3}(c). Next, we discuss the effect of these distortions in the magnetic order. 

The spin Hamiltonian can be written as $\mathcal H = \mathcal H_{\text{intra}}+\mathcal H_{\text{inter}}$, where the intralayer Hamiltonian has been proposed to correspond to a Heisenberg-Kitaev~\cite{kitaev2006,rau2014,Xu2018_cri3, xu2020,lee2020fundamental} system. Experimentally~\cite{chen2018,lee2020fundamental}, it has been determined that the intralayer exchange interaction are much stronger than the interlayer interactions. Furthermore, due to the inter-layer nature of the phonon modes here considered, the dominant effect is expected to manifest on the inter-layer sector of the Hamiltonian. We take into account up to third-nearest-neighbor exchange interactions $\mathcal H_{\text{inter}}=\frac{1}{2} \sum_{ij \in \text{int.}} J_{ij} \ss_i \cdot \ss_j$, and calculate $J_{ij}$ using a Green's function approach and the magnetic force theorem (for a detailed explanation , see Ref.~\cite{hoffmann2020}).  

The driven phonons effectively modify the distance between the Cr magnetic moments leading to a time-dependent exchange interaction~\cite{granado1999}
%
$
J[ \boldsymbol{u}(t) ] = J^0 +  \delta J \hat \dd \cdot   \boldsymbol{u}(t) + \mathcal O(\boldsymbol{u}(t)^2),
$
%
where $J^0$ is the equilibrium interaction, $\delta J$ is the strength of the first-order correction in the direction $\hat{\boldsymbol{ \delta}}$, and $\boldsymbol{u}(t)$ is the real-space phonon displacement. Since the inter-layer exchange interaction is the relevant energy scale of our problem, and it is much smaller than the frequency of the driven phonons, we use Floquet theory to find an effective interlayer exchange interaction of the form 
$J^{\text{eff}} = J^0 +  \delta J \hat{\boldsymbol{ \delta}} \cdot \langle \boldsymbol{u}_{\text{R}} \rangle,$
where  $\langle u_{\text{R}} \rangle$ is the time-averaged displacement. We calculate the effective spin interactions $J_{ij}$ as a function of the Raman displacement amplitude $Q_{\text{R}(2)}$. We define $J_{\perp} \equiv (1/2)\sum_{ij} J_{ij}$ such that  $J^{\text{eff}}_{\perp}(\langle Q_{\text{R}(2)} \rangle ) =J^0_{\perp} + \delta J_{\perp} \langle Q_{\text{R}(2)} \rangle$. We find $J^0_{\perp}= -0.366 $~meV and $\delta J_{\perp} =-0.0713$~meV/$(\text{\AA}\sqrt{\text{amu}})$, with $J^{\text{eff}}>0$, thus preferring FM order, for $\langle Q_{\text{R}(2)} \rangle < -5.13\text{\AA}\sqrt{\text{amu}}$ which corresponds to a real-space displacement of $\sim 3.13\%$ of the Cr-Cr interatomic distance. However, $J^0_{\perp}$ overestimates the experimental value for b-CrI$_3$~\cite{lee2020fundamental}. Using $J^0_{\perp}$ as a fitting parameter from experiments, and $\delta J_{\perp}$ from our calculations, we find $J^{\text{eff}}(\langle Q_{\text{R}(2)} \rangle)>0$ for $\langle Q_{\text{R}(2)} \rangle < -0.42\text{\AA}\sqrt{\text{amu}}$, $\sim0.3\%$ of the Cr-Cr interatomic distance.

Driving the phonons with low-frequency light can induce non-equilibrium lattice displacements with non-zero averages. For the case of b-CrI$_3$, we find that such distortions can induce an antiferromagnet to ferromagnet transition. Therefore, Floquet protocols no only allow for the modification of the quasienergy bands in the single-particle approximation but also can affect ordered states of matter. This protocols can be extended to the recently theoretically proposed moir\'e magnets \cite{Hejazi10721,PhysRevB.102.094404}.

\section{Summary and Perspectives}

In this review, we discussed recent developments in the theoretical methods available to derive effective Floquet Hamiltonians in the high-, mid-, and low-frequency regimes, with an emphasis on the latter.  We have described the advantages and disadvantages of each approach. These methods open the possibility of Floquet engineering of quantum materials.  In particular, the techniques are able to determine effective Floquet (time-independent) Hamiltonians in different regimes of the drive frequency.  We paid special attention to the flexibility and accuracy of these approaches to make various classes of approximations, depending on the details of the system of interest. 

We focused our attention on the application of these methods to moir\'e superlattices in van der Waals materials.  We applied these techniques to graphene-based bilayer and double bilayer materials, and other bilayer material classes exhibiting strong correlation effects, such as magnetism. In some classes of van der Waals materials, such as transition metal dichalcogenides, the spin-orbit coupling may be strong which also brings band topology into play.  The interaction of these materials with light may lead to control of the material's topological phase, or perhaps even the realization of new topological phases.  Since light can also flatten bands, tuning with light also offers an opportunity to change the relative strength of interactions in a material, as occurs with the twist around the ``magic angle" in graphene. 

Finally, it is worth emphasizing that we are just starting to unlock the low-frequency regime's potential. In this work we discussed the cases of driven phonons and driven electrons separately. However, when the symmetries are appropriate, the light pulses can couple to both the phonons and the electrons creating a coherent dance of interacting degrees of freedom that can lead to new  phases of matter providing yet another example where \textit{``More is different"}~\cite{Anderson393}.

\section{Aknowledgements}

We are grateful to Babak Seradjeh, Abhishek Kumar, Tami Pereg-Barnea, Meghan Lentz, Herbert A. Fertig, Fengchen Wu, Sam Shallcross, Gaurav Chaudhary, Arthur Ernst, Pontus Laurell, Aaron D. Barr,  Aritz Leonardo, Ze-Xun Lin, and Maia G. Vergniory, Edoardo Baldini, Carina A. Belvin, Ilkem Ozge Ozel, Dominik Legut, Andrzej Kozlowski, Andrzej M. Ole\'s, P. Piekarz, Jos\'e Lorenzana, Nuh Gedik, Benedetta Flebus, and Allan H. MacDonald for discussions and collaborations on some of the topics discussed here.  This research was primarily supported by the National Science Foundation through the Center for Dynamics and Control of Materials: an NSF MRSEC under Cooperative Agreement No. DMR-1720595.  We also acknowledge partial support under NSF DMR-1949701.


\end{document}